\documentclass[pdflatex,sn-mathphys-num]{sn-jnl}

\usepackage{graphicx}%
\usepackage{multirow}%
\usepackage{amsmath,amssymb,amsfonts}%
\usepackage{amsthm}%
\usepackage{mathrsfs}%
\usepackage[title]{appendix}%
\usepackage{xcolor}%
\usepackage{textcomp}%
\usepackage{manyfoot}%
\usepackage{booktabs}%
\usepackage{algorithm}%
\usepackage{algorithmicx}%
\usepackage{algpseudocode}%
\usepackage{listings}%

\usepackage{lipsum}
\usepackage{graphicx} 
\usepackage{amsmath}
\usepackage{amssymb}
\usepackage{bm}
\usepackage[dvipsnames]{xcolor}
\usepackage{cleveref}
\crefname{figure}{Fig.}{Figs.}
\usepackage{appendix}
\usepackage{booktabs}
\usepackage{multirow}
\usepackage{enumitem}
\usepackage{array}

\newcommand{\mU}{\mathcal{U}} 
\newcommand{\mN}{\mathcal{N}} 

\theoremstyle{thmstyleone}%
\theoremstyle{thmstyletwo}%

\theoremstyle{thmstylethree}%

\makeatletter
\def\ps@preprint{%
  \ps@headings
  \def\@oddfoot{%
    \parbox[t]{\textwidth}{\raggedright\fontsize{8}{10}\selectfont
      \rule{0.4\textwidth}{0.4pt}\par\vspace{4pt}
      This arXiv version is a preprint. Accepted for publication in \textit{PNAS}.\\
      Final published version in \textit{PNAS}: \href{https://doi.org/10.1073/pnas.2620983123}{doi: 10.1073/pnas.2620983123}.\par
      \vspace{8pt}{\centering\normalsize\thepage\par}}}%
  \let\@evenfoot\@oddfoot}
\makeatother

\begin{document}

\title{In-context learning as a retraining-free paradigm for constitutive modeling}

\author[1,2]{\fnm{Lingfeng} \sur{Li}}
\equalcont{These authors contributed equally to this work.}

\author[4,5]{\fnm{Zhuoyuan} \sur{Li}}
\equalcont{These authors contributed equally to this work.}

\author[1]{\fnm{Shun} \sur{Li}}
\equalcont{These authors contributed equally to this work.}

\author[2]{\fnm{Kaixin} \sur{Zhan}}

\author*[3]{\fnm{Huajian} \sur{Gao}}\email{gao.huajian@tsinghua.edu.cn}

\author*[1,3]{\fnm{Changqing} \sur{Chen}}\email{chencq@tsinghua.edu.cn}

\author*[2]{\fnm{Liu} \sur{Yang}}\email{yangliu@nus.edu.sg}

\affil[1]{\orgdiv{Department of Engineering Mechanics}, \orgname{Tsinghua University}}

\affil[2]{\orgdiv{Department of Mathematics}, \orgname{National University of Singapore}}

\affil[3]{\orgdiv{Mechano-X Institute}, \orgname{Tsinghua University}}

\affil[4]{\orgdiv{Institute for Functional Intelligent Materials}, \orgname{National University of Singapore}}

\affil[5]{\orgdiv{School of Mathematical Sciences}, \orgname{Peking University}}

 
\abstract{

Building models that generalize across physical systems without retraining remains a central challenge in computational science. Here we introduce In-Context Modeling (ICM), a retrain-free paradigm that infers physical relationships directly from observational fields. Rather than encoding system-specific behavior in fixed parameters, ICM assimilates measurements as physical context and performs inference through a single forward pass. Trained in a physics-informed, label-free manner using governing equations, a single model generalizes across unseen materials, geometries, and loading conditions. Demonstrated on hyperelasticity, ICM integrates with finite-element simulations and is validated using experimental full-field measurements. Moreover, performance improves with increasing data diversity and computational budget, exhibiting favorable scaling behavior analogous to foundation models. By recasting physical modeling as in-context inference, this work establishes a transferable paradigm for retrain-free scientific learning and a foundation for scalable modeling across computational science.
}





\thispagestyle{preprint}
\maketitle

Building models that generalize across physical systems without retraining remains a central challenge in computational science~\cite{wang2023scientific}. While modern data-driven methods achieve high accuracy for specific tasks~\cite{brunton2016discovering,rudy2017data,chen2021physics,yu2025discover}, they typically require retraining once the underlying physical relationship changes, leading to substantial computational overhead and limiting deployment in dynamic environments. Developing models that can adapt instantly to new scenarios is therefore a fundamental goal in scientific machine learning.

We argue that three fundamental bottlenecks hinder such generalization: the prevailing modeling-by-optimization paradigm, inadequate exploitation of governing physics, and an inability to scale with data and computational resources. Existing approaches encode physical relationships into static model parameters through case-specific optimization, making adaptation to new systems inherently expensive. Moreover, failure to isolate intrinsic physical structure from extrinsic factors often leads to spurious correlations, further limiting generalization.

Here we introduce In-Context Modeling (ICM) as a paradigm that overcomes these limitations by shifting from modeling-by-optimization to modeling-from-context. In this paradigm, a single model infers physical relationships by conditioning on observational data treated as physical context, without parameter updates. Drawing inspiration from in-context learning in large language models, ICM enables retrain-free adaptation across diverse physical systems (\cref{fig:1}a).

Grounded in a common mathematical structure shared across physical disciplines, ICM integrates physics at multiple levels, including tokenization of governing equations, symmetry-preserving architectures, and label-free training based on physical laws~\cite{karniadakis2021physics}. As a result, ICM generalizes across materials, geometries, and loading conditions, while exhibiting favorable scaling behavior with increasing data diversity and computational budget.

We demonstrate these capabilities on the challenging problem of constitutive modeling in hyperelasticity~\cite{rivlin1951large,treloar1975,xu2021learning,flaschel2021unsupervised,li2022equilibrium,thakolkaran2022nn,linden2023neural}, where ICM infers stress-strain relationships directly from deformation fields and generalizes to unseen scenarios without retraining. By recasting physical modeling as in-context inference, this work establishes a transferable paradigm for retrain-free scientific learning across computational mechanics and beyond.

\section*{Results}
\subsection*{In-Context Modeling}\label{sec:in-context modeling}

\begin{figure}
    \centering
    \includegraphics[width=0.98\textwidth]{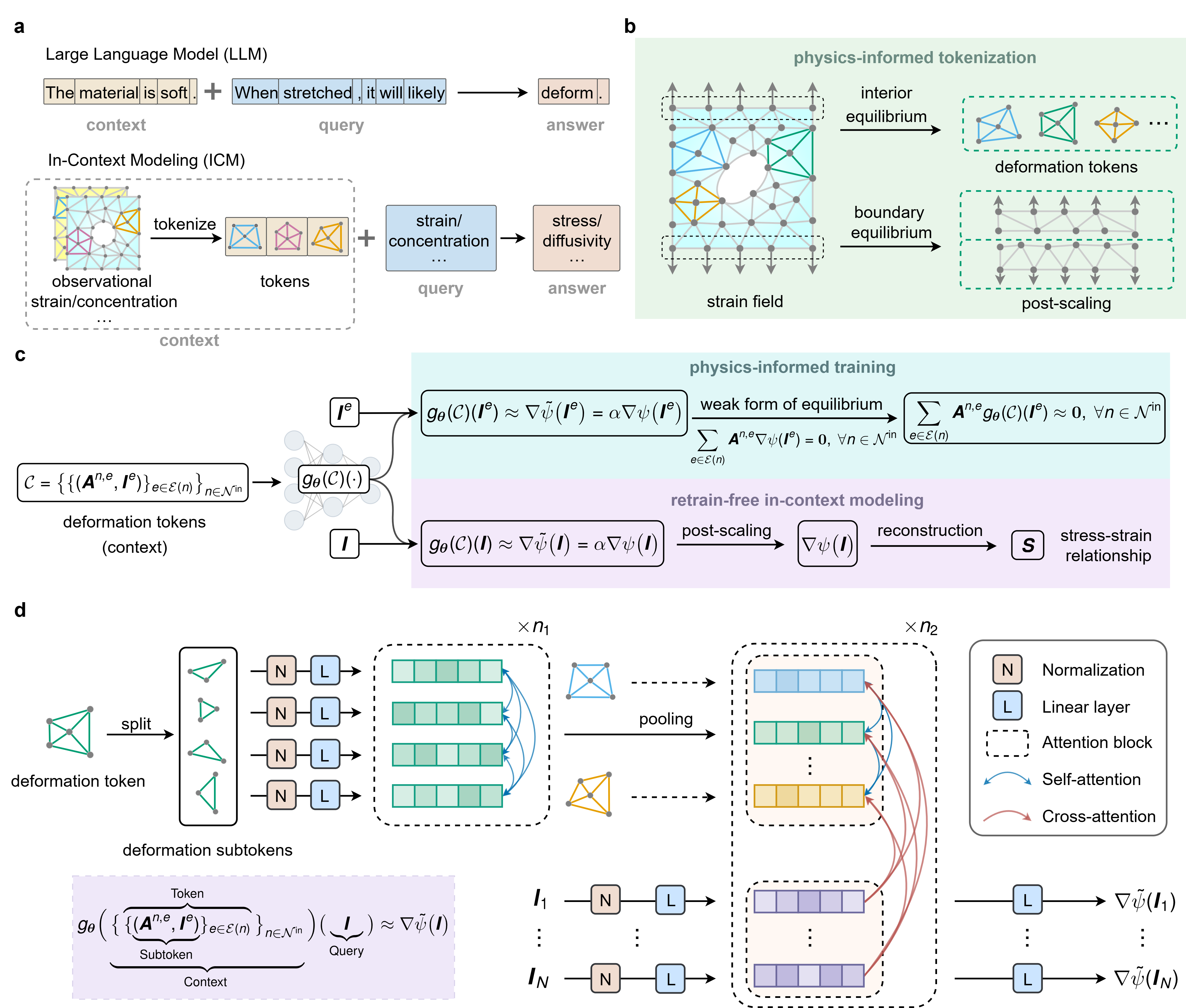}
    \caption{\textbf{Schematic overview of the ICM paradigm.} 
    \textbf{a,} Conceptual analogy between a Large Language Model (LLM) and the proposed In-Context Modeling (ICM). The LLM learns to infer an answer given its textual context; ICM learns to infer the stress-strain or diffusivity-concentration relationship given its physical context.
    \textbf{b,} Physics-informed tokenization in ICM. The strain field is discretized into deformation tokens based on interior equilibrium, while the boundary equilibrium is extracted for the purpose of post-scaling.
    \textbf{c,} The overall workflow of ICM. Assimilating deformation tokens as the physical context, the core network $g_{\bm{\theta}}$ predicts the scaled strain energy gradient function. In the physics-informed training phase, we enforce interior equilibrium in an unsupervised manner. In the retrain-free in-context modeling phase, we recover the stress-strain relationship via a single forward pass, followed by lightweight post-scaling and analytical reconstruction.
    \textbf{d,} The network architecture of the ICM network $g_{\bm{\theta}}$. 
    We first map each deformation token into a context embedding by applying normalization, linear projection, self-attention, and average pooling to its subtokens. Concurrently, input strain queries are independently encoded into query embeddings via normalization and linear projection. These context and query embeddings then interact via self- and cross-attention mechanisms to exchange and aggregate information, before a final linear projection which yields the network output.
    }
    \label{fig:1}
\end{figure}

To introduce In-Context Modeling, we first discuss the mathematical formulation of physical modeling, which in general refers to the task of inferring the unknown physical relationship under which the observed behaviors satisfy a universal physical law, such as equilibrium and conservation. As we will show through concrete examples, diverse physical disciplines usually share a common mathematical structure upon discretization and reformulation:
\begin{equation}\label{eq:universal}
    \sum_{e\in\mathcal E(n)}
    \bm A^{n,e} \bm{f}(\bm x^{n,e}) = \bm b^{n},
    \qquad
    \forall n\in\mathcal N.
\end{equation}
Here, $\mathcal N$ denotes the set of equations, and $\mathcal E(n)$ denotes the set of components $e$ in the $n$-th equation. In this formulation, the known coefficient matrices $\bm A^{n,e}$ and the vectors $\bm x^{n,e}$, $\bm b^{n}$ encapsulate the observations, $\bm f(\bm x)$ represents the unknown relationship, and the equality constraint arises from the governing universal law. The challenge of physical modeling thus translates into inferring $\bm f(\bm x)$ given $\bm A^{n,e}$, $\bm x^{n,e}$, and $\bm b^{n}$.

As an example, we consider the task of hyperelastic modeling. Given a set of observed equilibrium strain fields and boundary conditions for an unknown isotropic hyperelastic material, we aim to infer the stress-strain relationship. 

Without loss of generality, we consider the following weak formulation of force equilibrium~\cite{de2012nonlinear,belytschko2014nonlinear}:
\begin{equation}\label{eq:weak-form-simple}
    \int_{\Omega} \bm P : \nabla \bm v~\mathrm{d}V = 0, \quad  \forall \bm v\in\mathcal V,
\end{equation}
where $\Omega$ denotes the domain, $\bm P$ is the first Piola-Kirchhoff stress tensor, $\bm v$ is the test function in space $\mathcal V$.

By applying discretization and tensor transformations (see `Derivation of modeling objective in hyperelasticity' in Methods), \cref{eq:weak-form-simple} yields
\begin{equation}\label{eq:equilibrium-internal}
    \sum_{e\in\mathcal{E}(n)}\bm A^{n,e} \nabla \psi (\bm I^e)=\bm 0, \quad \forall n \in \mathcal{N}^{\rm in},
\end{equation}
where $\mathcal{N}^{\rm in}$ represents the aggregated set of all interior nodes collected from strain fields, $\mathcal E(n)$ denotes the set of elements connected to node $n$, $\bm I^e$ is the strain invariants of element $e$, $\psi$ is the strain energy density function, and $\bm A^{n,e}$ is the coefficient matrix associated with $n$ and $e$, determined by the local deformation and mesh configuration. For isotropic hyperelastic materials, $\psi$ is a function of $\bm I$, and we use $\nabla \psi$ to denote its gradient with respect to input $\bm I$. Physically, each equation establishes the nodal force equilibrium for an interior node, with the summation terms accounting for the contributions from adjacent elements. 

Note that \cref{eq:equilibrium-internal} is indeed a special form of \cref{eq:universal} by viewing $\nabla \psi$ as the unknown function $\bm f$. This is not an isolated case. Let's consider a distinct task: inferring the relationship between diffusivity tensor $\bm D$ and concentration $c$ from an diffusion process governed by mass conservation:
\begin{equation}\label{eq:diffusion-weak-main}
    \int_{\Omega} v \frac{\partial c}{\partial t}~\mathrm dV + \int_{\Omega} \nabla v\cdot\big(\bm D(c)\nabla c\big)~\mathrm dV = 0, \qquad \forall v\in\mathcal V.
\end{equation}
By applying spatial and temporal discretization, \cref{eq:diffusion-weak-main} elegantly reduces to
\begin{equation}\label{eq:diffusion-internal}
    \sum_{e\in\mathcal E(n)}
    \bm A^{n,e,m} : \bm D(c^{e,m}) = b^{n,m},
    \qquad
    \forall n\in\mathcal N^{\rm in},\ \forall m\ge 1,
\end{equation}
where $m$ denotes the time step, $\bm A^{n,e,m}$ is derived from the local concentration gradients and mesh configurations, and $b^{n,m}$ accounts for the transient variation of concentration (see `Derivation of modeling objective in nonlinear diffusion' in Methods). Once again, \cref{eq:diffusion-internal} aligns with \cref{eq:universal} if we view $\bm D$ as the unknown relationship $\bm f$. 

Despite originating from entirely different physics, both hyperelasticity and nonlinear diffusion ultimately converge to a common mathematical structure. While the remainder of this study focuses on hyperelastic modeling, ICM built upon such a mathematical structure can seamlessly adapt to other disciplines.

Returning to \cref{eq:equilibrium-internal}, we can aggregate the set $\{(\bm A^{n,e}, \bm I^e)\}_{e\in \mathcal{E}(n)}$ into a ``deformation token'', where each individual term $(\bm A^{n,e}, \bm I^e)$ acts as a ``deformation subtoken'' (\cref{fig:1}b). Such physics-informed tokenization translates the governing equilibrium into a structured representation. By encapsulating the local equilibrium surrounding an interior node and stripping away ad hoc global domain geometry and loading conditions, each token acts as the atomic unit of information regarding the stress-strain relationship. This inherent modularity allows us to freely sample and aggregate tokens from an arbitrary number of strain fields (see `Network inputs and architecture' in Methods). When these tokens are assembled to form the physical context $\mathcal{C} = \{\{(\bm A^{n,e}, \bm I^e)\}_{e\in \mathcal{E}(n)}\}_{n \in \mathcal{N}^{\rm in}}$, they provide a comprehensive representation of the interior equilibrium across strain fields.

The core of ICM is a neural network $g_{\bm\theta}$ with trainable parameters $\bm\theta$, which approximates $\nabla \psi$ given the context $\mathcal{C}$, up to a scaling factor (\cref{fig:1}c,d). By mapping the physical context directly to the scaled energy density gradient instead of specific material parameters, ICM is inherently agnostic to the mathematical form of the strain energy, naturally generalizing to unseen material classes. Formally, the network treats the deformation tokens as the physical context, takes the strain invariants $\bm I$ as the query, and predicts the corresponding scaled energy gradient at the queried $\bm I$, i.e.,
\begin{equation}\label{eq:g-theta}
    g_{\bm\theta} \Big( 
    \underbrace{ 
        \big\{ 
        \overbrace{ 
            \{ \underbrace{(\bm A^{n,e}, \bm I^e)}_{\text{Subtoken}} \}_{e\in \mathcal{E}(n)} 
        }^{\text{Token}} 
        \big\}_{n \in \mathcal{N}^{\rm in}} 
    }_{\text{Context }} 
    \Big) 
    \big( \underbrace{ \bm I }_{\text{Query}} \big) 
    \approx \nabla \tilde{\psi} \big( \bm I \big) = \alpha \nabla \psi \big( \bm I \big),
\end{equation}
where $\alpha$ is the scaling factor that can be calculated during inference. To address the spatial heterogeneity in mesh resolution and the variation in strain ranges, we normalize all deformation tokens and queries utilizing the scale invariance of \cref{eq:equilibrium-internal}. This preprocessing step ensures that network inputs remain invariant to non-physical discretization artifacts and are adaptively scaled to the contextual deformation state. For the network architecture, the ICM network adopts self- and cross-attention mechanisms~\cite{vaswani2017attention}, which not only generalize to a variable number of deformation tokens and subtokens, but also maintain their permutation invariance, respecting the unordered nature of the nodal equilibrium equations (see `Network inputs and architecture' in Methods).

By substituting the neural approximation \cref{eq:g-theta} into the discretized equilibrium equations \cref{eq:equilibrium-internal}, we obtain
\begin{equation}\label{eq:neural-equilibrium}
    \sum_{e\in\mathcal{E}(n)} \bm A^{n,e} g_{\bm\theta} ( 
        \mathcal{C}
    ) 
    ( \bm I^e ) 
    \approx \bm 0, \quad \forall n \in \mathcal{N}^{\rm in}.
\end{equation}
The ICM network is trained by minimizing the residuals of \cref{eq:neural-equilibrium} as a physics-informed loss that enforces discretized equilibrium. 
This formulation enables unsupervised learning without the need for ground-truth stress information.

As \cref{eq:equilibrium-internal} is invariant to the scaling of the network prediction, an additional post-scaling process is required to determine the scaling factor $\alpha$ during inference. To this end, we revisit the weak form of equilibrium \cref{eq:weak-form-simple} on the boundaries, which implies
\begin{equation}\label{eq:equilibrium-external}
    \sum_{n\in \mathcal{N}^{\rm b}_{j,k}}\sum_{e\in\mathcal{E}(n)}\bm A^{n,e}\nabla \psi (\bm I^e)=\bm F_{j,k}, \quad \forall (j,k) \in \mathcal{B},
\end{equation}
where $\mathcal{N}^{\rm b}_{j,k}$ indicates all boundary nodes under the $j$-th boundary condition of the $k$-th contextual strain field, $\bm F_{j,k}$ is the corresponding external force, and $\mathcal{B}$ denotes the index set of all $(j,k)$ pairs. To obtain the scaling factor $\alpha$, we can compute $\alpha_{j,k}$ for each $(j,k)$ pair by comparing the true external force magnitude with the inferred one calculated by replacing $\nabla \psi$ with $g_{\bm\theta}$ in \cref{eq:equilibrium-external}, and then average $\alpha_{j,k}$ over $\mathcal{B}$ as the estimated $\alpha$. Once $\nabla \psi$ is obtained, the stress $\bm S$ is analytically reconstructible. We refer to this rescaling as the post-scaling procedure (see `Training and inference' in Methods). This scale-agnostic training and post-scaling approach is critical to the model's generalization across materials: ICM decouples the profile of the stress-strain relationship from its absolute magnitude, allowing the network to remain robust even for materials exhibiting orders-of-magnitude variations in stress.

\subsection*{Dataset and evaluation setup}

\begin{figure}
    \centering
    \includegraphics[width=1.0\textwidth]{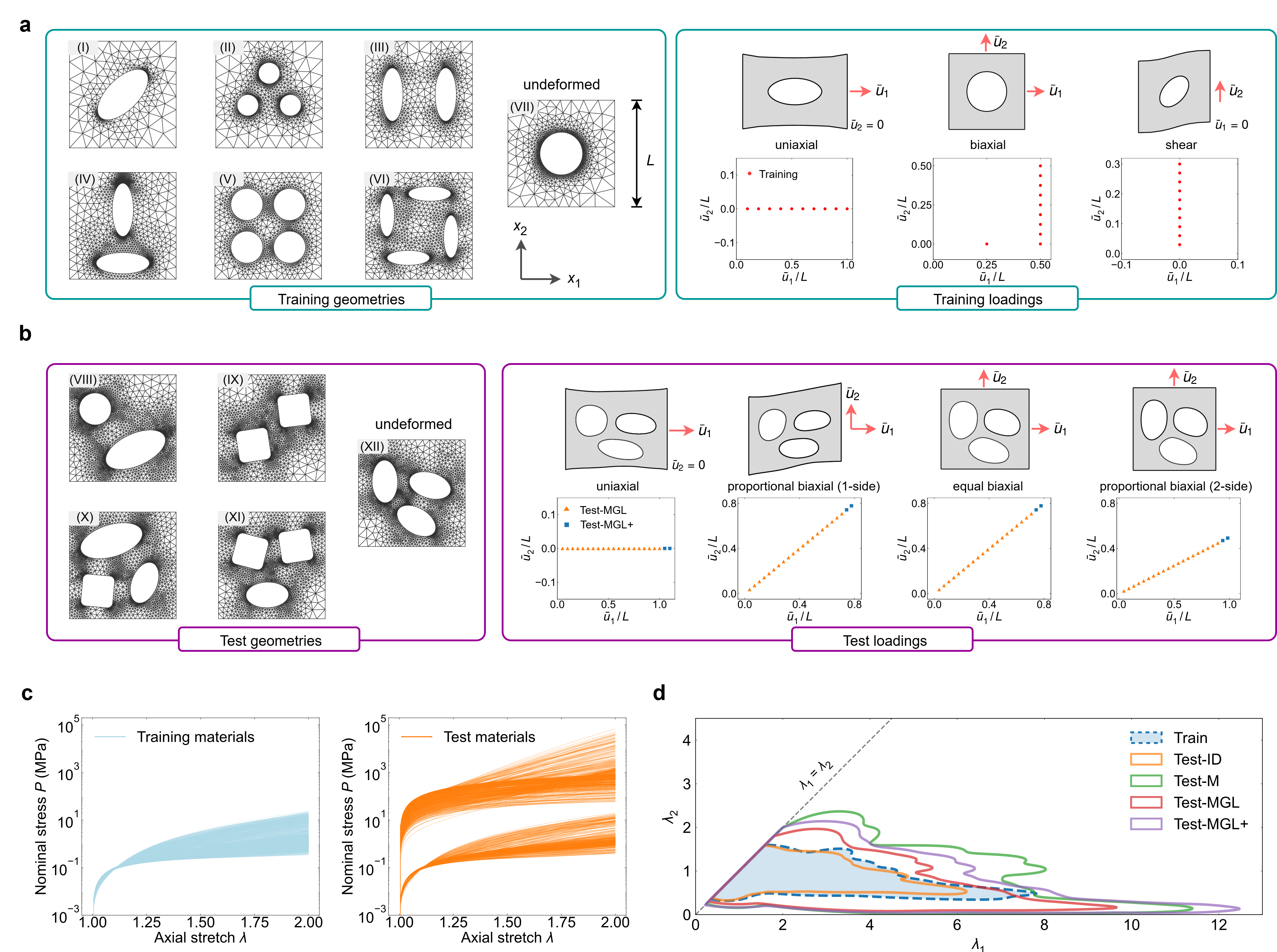}
    \caption{
    \textbf{Configurations of the training and test sets.}
    \textbf{a,} Geometries and loading modes in the training dataset, including 7 plate geometries with different hole shapes and arrangements (labeled I-VII), 3 loading modes, and 10 loading steps for each loading mode.
    \textbf{b,} Geometries (labeled VIII--XII) and loading modes in the Test-MGL and Test-MGL+ sets. Note the inclusion of unseen square-hole geometries and new loading modes not present in (\textbf{a}).
    \textbf{c,} Material differences between training (left) and test (right) sets, shown by uniaxial nominal stress-stretch ($P$-$\lambda$) curves, highlighting orders-of-magnitude variations in stress levels.
    \textbf{d,} Distribution of deformation states in training and test sets, shown in the principal stretch space ($\lambda_{1}$-$\lambda_{2}$). The contours represent the effective coverage of each dataset, showing that the test sets span regions extending beyond the training domain, thereby providing a comprehensive testbed for evaluating the model's out-of-distribution generalization.
    }
    \label{fig:2}
\end{figure}

For a training dataset covering diverse scenarios, we generated 2,000 hyperelastic material models defined by polynomial strain energy functions (Supplementary Section 1) and combined them with seven plate geometries with different numbers and arrangements of circular or elliptical holes (\cref{fig:2}a). Each geometry is subjected to uniaxial tension, biaxial tension, and in-plane shear, producing complex, non-uniform strain fields. Altogether, the dataset contains over 500 million (M) deformation tokens and captures a broad spectrum of hyperelastic mechanical responses.

To verify robust generalization, we construct four test sets that progressively increase the data diversity: (i) \textbf{Test-ID} evaluates in-distribution prediction by replacing the training materials with 400 unseen polynomial hyperelastic models while keeping the training geometries and loading modes unchanged. (ii) \textbf{Test-M} introduces 500 unseen materials drawn from four other common strain energy forms, namely Ogden~\cite{ogden1972large,ogden1997non}, Pucci-Saccomandi (PS)~\cite{pucci2002note}, Exp-ln~\cite{khajehsaeid2013hyperelastic}, and van der Waals (VdW)~\cite{kilian1986use,marckmann2006comparison} (Supplementary Section 1.1), again under the training geometries and loading modes. (iii) \textbf{Test-MGL} uses the same materials as Test-M, but associates them with five unseen geometries and additional loading modes (\cref{fig:2}b). (iv) \textbf{Test-MGL+} further increases max loading magnitudes in Test-MGL by 10\%. This incremental setup provides a multifaceted lens for examining how the model generalizes across increasingly complex unseen scenarios, spanning from new materials to unexplored geometric and loading configurations.

Material differences between training and test sets are summarized in \cref{fig:2}c, highlighting the orders-of-magnitude variations in stress levels between them. \cref{fig:2}d compares the distribution of deformation states in each dataset. Notably, the test sets span regions extending beyond the training domain, reflecting the compounded effects of distribution shifts in materials, geometries, and loadings, thereby providing a comprehensive testbed for evaluating the model's out-of-distribution generalization.

\subsection*{Generalization in materials, geometries, and loadings}

\begin{figure}
    \centering
    \includegraphics[width=1.0\textwidth]{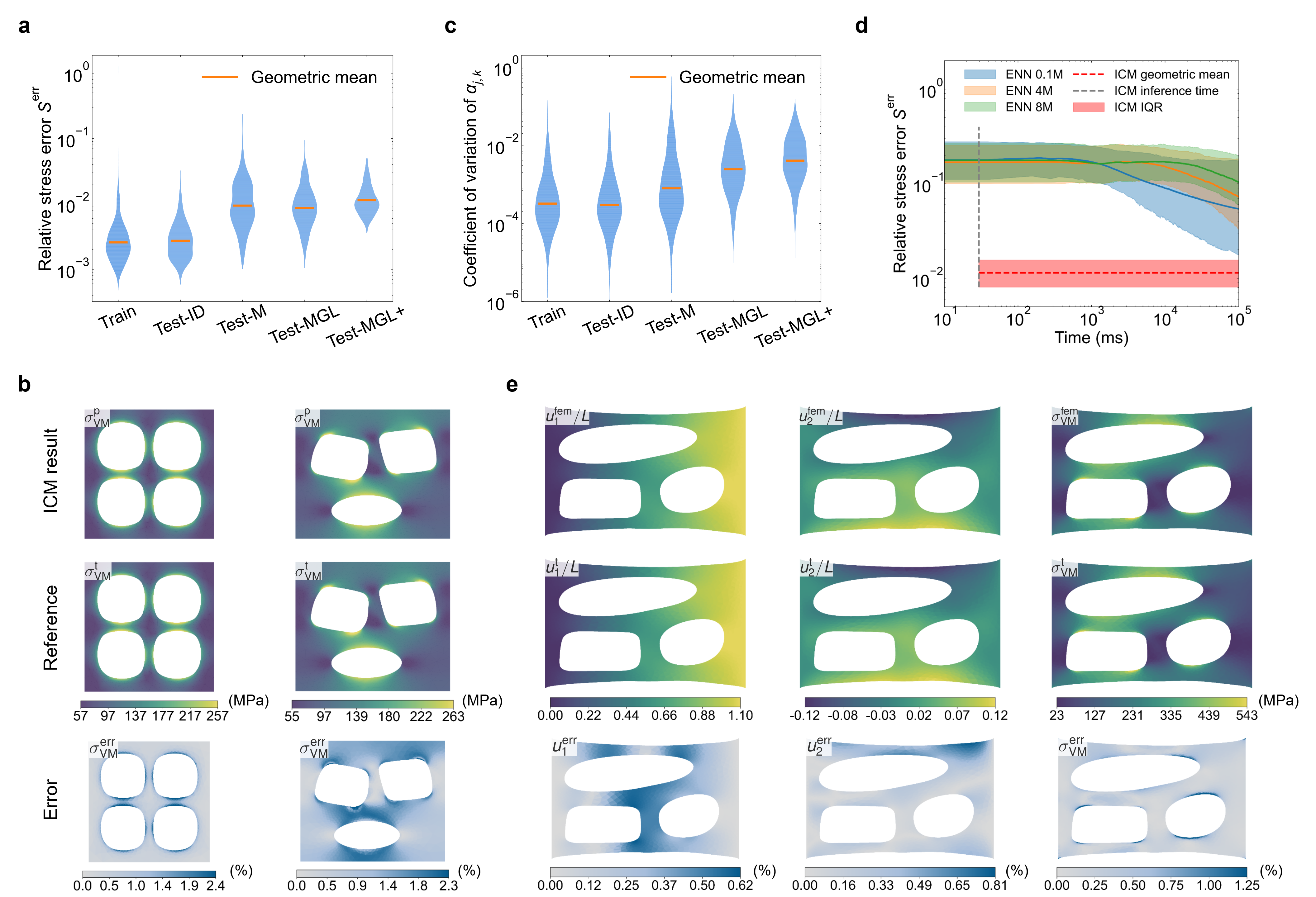}
    \caption{
    \textbf{Generalization performance and finite-element validation of ICM.}
    \textbf{a,} Relative stress error $S^{\mathrm{err}}$ on the training set and the four test sets, where the blue shaded areas and orange lines denote the underlying distributions and geometric means across samples, respectively. The relative errors remain in the order of $10^{-2}$ for all datasets, showing robust generalization.
    \textbf{b,} Visualization of von Mises stress fields for two representative samples from Test-M (left) and Test-MGL$+$ (right). Each column shows the prediction ($\sigma_{\mathrm{VM}}^{\mathrm{p}}$), ground truth reference ($\sigma_{\mathrm{VM}}^{\mathrm{t}}$), and spatial distribution of relative error ($\sigma_{\mathrm{VM}}^{\mathrm{err}}$).
    \textbf{c,} Coefficients of variation for the scaling factors $\alpha_{j,k}$ on the training set and the four test sets, confirming consistency among different boundary conditions.
    \textbf{d,} Comparison between ICM with 4\,M parameters and the Equilibrium-based Neural Network (ENN) with 0.1\,M, 4\,M, and 8\,M parameters. ICM achieves lower errors while reducing computation time by more than three orders of magnitude. The solid lines and shaded regions represent the geometric mean and interquartile range (IQR) across test materials, respectively.
    \textbf{e,} ICM-driven finite element simulations in FEniCS for a Test-MGL$+$ sample under uniaxial displacement loading ($\bar{u}_{1}/L=1.1$). Predicted displacement components ($u_{1}^{\mathrm{fem}}/L$ and $u_{2}^{\mathrm{fem}}/L$, visualized with a scale factor of 0.5), and von Mises stress ($\sigma_{\mathrm{VM}}^{\mathrm{fem}}$) closely match the reference finite element solution.}
    \label{fig:3}
\end{figure}

With strategic designs in tokenization, preprocessing and network architecture, as well as the training scheme, we expect to equip ICM with a generalizable capability to infer intrinsic stress-strain relationships from physical context. To validate this expectation, we train an ICM network with 4\,M parameters for $3.4 \times 10^5$ steps, and look into the relative prediction error $(\cdot)^\mathrm{err}$ (see `Training and inference' in Methods), treating each individual tokenized deformation field from test sets as both context and query.

\cref{fig:3}a summarizes the results across the training and test sets, with representative examples visualized in \cref{fig:3}b. Even for the most challenging Test-MGL+ dataset, the relative errors remain in the order of $10^{-2}$, showing robust generalization. Interestingly, the prediction error for Test-MGL is slightly lower than that for Test-M. This counter-intuitive result is, in fact, a direct manifestation of our physics-informed design. As revealed in \cref{fig:2}d, despite the introduction of unseen geometries and loading modes, the local deformation states in Test-MGL are actually closer to the training domain than those in Test-M. Since our method relies exclusively on these local states via the tokenization strategy, the lower prediction error in Test-MGL naturally follows.

For the post-scaling step, the final scaling factor is obtained by averaging the individual factors $\alpha_{j,k}$ derived from each boundary condition. The minimal coefficients of variation (standard deviation divided by the mean) observed across all datasets in \cref{fig:3}c confirm strong agreement among the factors inferred from different boundary conditions.

Existing equilibrium-based unsupervised stress inference methods are typically trained for each material without cross-material generalization~\cite{xu2021learning,li2022equilibrium,thakolkaran2022nn, li2025ennstressnet}. We compared ICM with the Equilibrium-based Neural Network (ENN)~\cite{li2022equilibrium}, a representative baseline for unsupervised stress inference (Supplementary Section 2). This comparison is appropriate because both methods are trained using only interior force equilibrium and do not require stress labels. Since ENN is trained in a material-specific way and does not generalize across materials, we trained ENN on the Test-M set and evaluated it on the Test-MGL+ set, which contains the same materials but introduces unseen geometries and loadings. In contrast, the ICM network was trained on its original training set and applied directly to Test-MGL+ without retraining, reflecting its ability to generalize to unseen materials. We used a 4\,M-parameter ICM network and compared it against ENN models of three sizes: 0.1\,M, 4\,M, and 8\,M. Deployment costs for new materials are measured by the total wall-clock time for stress predictions: for ENN, this comprises Test-M training plus Test-MGL+ inference, while for ICM, it includes only inference as no retraining is required. As shown in \cref{fig:3}d, ICM achieves lower relative stress errors while reducing the overall deployment time by more than three orders of magnitude compared with all ENN variants.

To demonstrate practical applicability, we integrated the trained ICM network as a constitutive model into the open-source finite element software FEniCS~\cite{AlnaesEtal2015,LoggEtal2012} (\cref{fig:3}e). In this pipeline, ICM first infers the material's stress-strain relationship from a provided context of 10 deformation fields, which is then utilized in FEniCS to solve a displacement loading problem. Notably, the geometry and loading configurations in the FEM simulation are entirely distinct from those in the ICM input context. The resulting displacement and von Mises stress fields show excellent agreement with ground truth, confirming that the stress-strain relationship inferred by ICM is fully compatible with conventional FEM pipelines.

\subsection*{Experimental validation and test-time scaling behavior}

\begin{figure}
    \centering
    \includegraphics[width=1.0\textwidth]{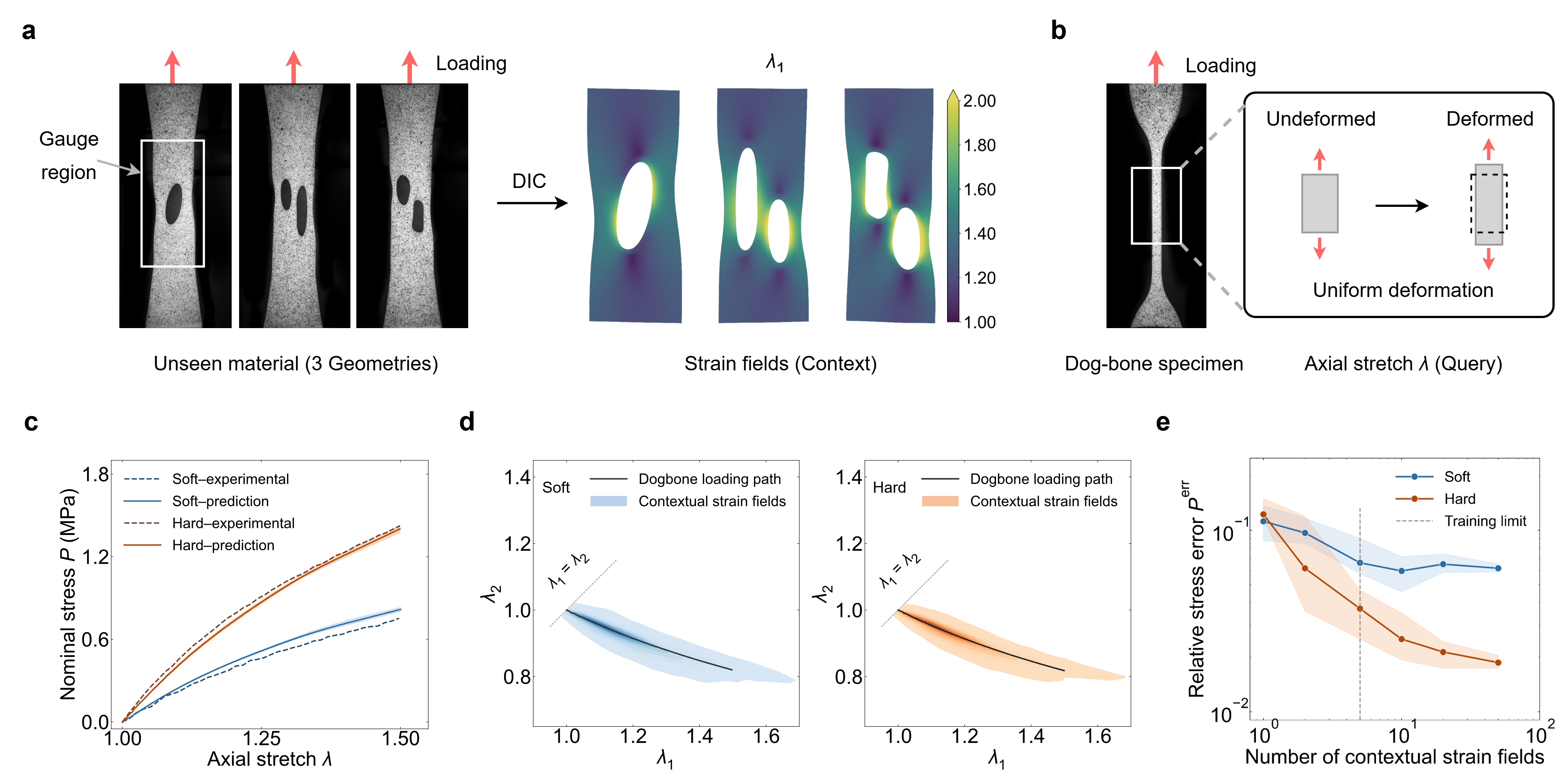}
    \caption{\textbf{Experimental validation of ICM.}
    \textbf{a,} Non-uniform strain fields were captured by applying DIC to 3D-printed perforated specimens under uniaxial tension. These fields, visualized by the maximum principal stretch $\lambda_1$, serve as the physical context for the network after tokenization.
    \textbf{b,} The uniform deformation, induced by independent uniaxial tests on standard dog-bone specimens and measured via DIC, is extracted as the query state.
    \textbf{c,} Comparison between the ICM-inferred nominal stress-stretch ($P$-$\lambda$) curves and the experimental ground truth. The predictions show excellent agreement for both materials without any material-specific retraining.
    \textbf{d,} Deformation state distributions in the principal stretch ($\lambda_1$-$\lambda_2$) space. The non-uniform contextual strain fields (shading indicates data density) provide broad coverage that fully envelopes the 1D strain trajectory of the dog-bone query, driving the accurate stress predictions in (\textbf{c}).
    \textbf{e,} Test-time scaling and context length extrapolation. The geometric mean (solid lines) and interquartile range (IQR) (shaded regions) of the relative nominal stress error $P^\mathrm{err}$ across random context samplings both decay with richer contextual strain fields. Notably, the sustained error reduction far beyond the training context length limit (vertical dashed line) demonstrates a remarkable capability for context length extrapolation at inference.
    }
    \label{fig:4}
\end{figure}  

To test whether ICM can infer accurate stress-strain relationships from experimentally measurable quantities, specifically the full-field strain measurements from Digital Image Correlation (DIC)~\cite{chu1985applications,bay1999digital,pan2009two,wang20243d,buljac2018digital} and the external forces recorded by the testing machine, we validated ICM via real-world experiments on two hyperelastic materials, classified as ``soft'' and ``hard'' based on their relative stiffness (Supplementary Section 3). 

For each material, 3D-printed perforated specimens with distinct geometries were tested under uniaxial tension to induce non-uniform deformation fields (\cref{fig:4}a), which were recorded via DIC to serve as context inputs for ICM. Given the absence of ground-truth stress-strain relationships, direct stress comparisons are infeasible. To establish a quantitative comparison, a standard dog-bone specimen of the same material was independently tested under uniaxial tension (\cref{fig:4}b). This setup yielded a uniform deformation field measured by DIC, alongside nominal stress derived directly from the external force. By utilizing tokenized non-uniform fields as context and the uniform dog-bone strain as the query, the ICM-predicted stresses exhibit excellent agreement with the experimental ground truth for both materials (\cref{fig:4}c). To explain this performance, \cref{fig:4}d visualizes the contextual deformation states, which broadly cover the entire strain trajectory of the dog-bone query, thereby enabling the highly accurate stress predictions.

In modern Large Language Models, the emerging paradigm of ``test-time scaling'' demonstrates that a model's performance can consistently improve by allocating additional resources during inference, such as processing richer contextual information. ICM exhibits an analogous behavior. As shown in \cref{fig:4}e, both the prediction error and its interquartile range decrease with more context. Notably, this scaling behavior shows a remarkable capacity for context length extrapolation. While the network was trained with a limited range of up to 1500 tokens sampled from a maximum of 5 strain fields, it successfully extrapolates to an order of $5\times 10^4$ tokens from 50 strain fields during inference with continual performance gains. 

\subsection*{Learning the intrinsic manifold}

\begin{figure}
    \centering
    \includegraphics[width=1.0\textwidth]{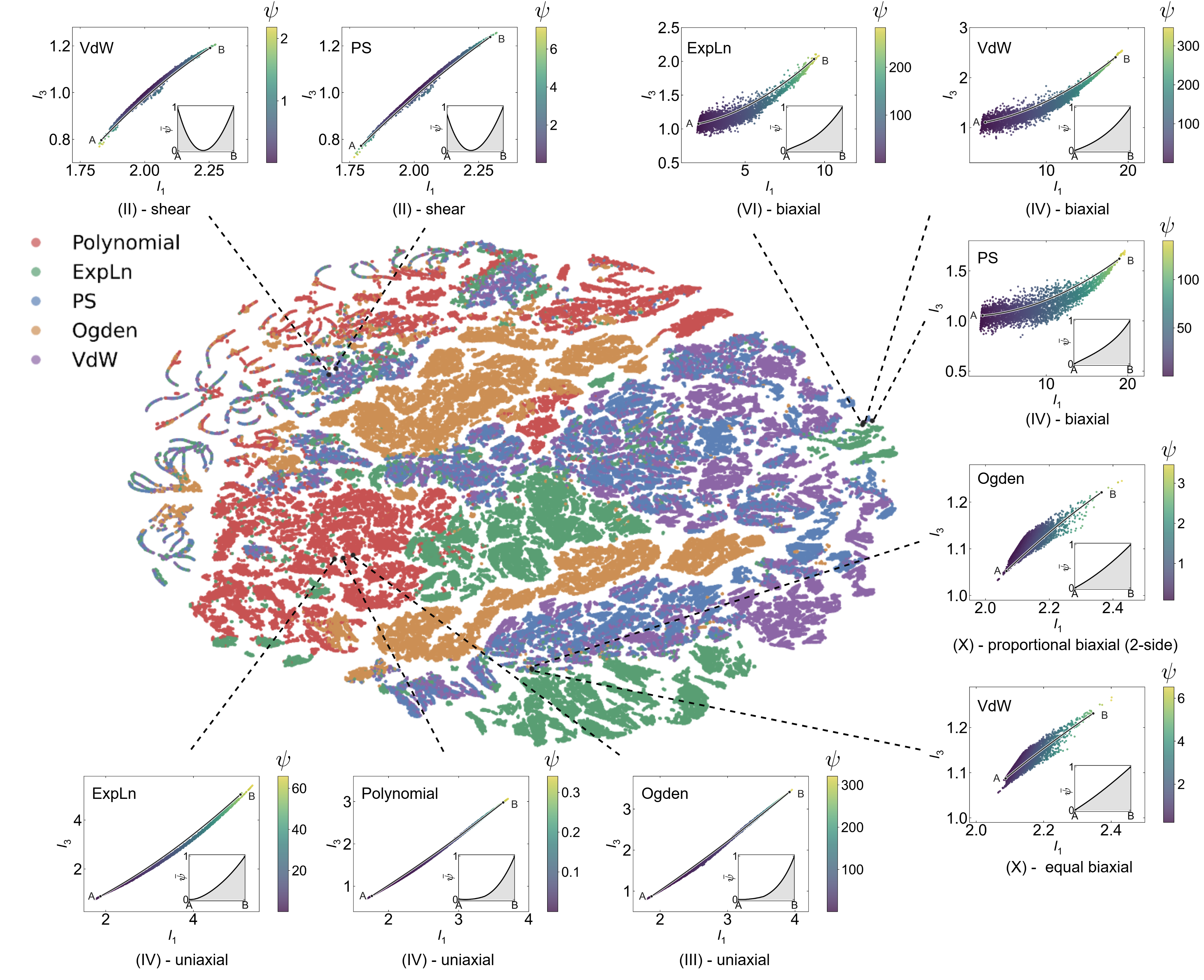}
    \caption{
    \textbf{Visualization of the learned context embedding manifold.} 
    We projected the average-pooled context embeddings of diverse deformation fields into a two-dimensional t-SNE space. The insets visualize the underlying strain energy density $\psi$ restricted to $(I_1, I_3)$ states within each deformation field, labeled with their material classes, geometries, and loading modes. In each inset, a curve from A to B marks the selected deformation path, and the smaller panel shows the corresponding normalized energy $\bar{\psi}$ along the path. Crucially, adjacent points in the manifold correspond to nearly identical strain energy density profiles, despite highly diverse material classes, energy magnitudes, geometries, and loading conditions. This visual evidence demonstrates that ICM successfully learns an intrinsic manifold determined primarily by the profiles of stress-strain relationships rather than other ad hoc factors, thereby driving its robust empirical generalization.
    }
    \label{fig:5}
\end{figure}

Having demonstrated the robust empirical generalization of ICM across unseen testing scenarios, we now investigate the driving mechanisms. To this end, we analyzed the model's latent representations across all test sets, encompassing 5 material classes, 12 geometries, and 7 loading modes. Specifically, we visualized the average-pooled final-layer context embeddings of each deformation field using the t-distributed Stochastic Neighbor Embedding (t-SNE) technique~\cite{maaten2008visualizing} (\cref{fig:5}). The t-SNE manifold preserves local neighborhood relationships, allowing us to inspect how ICM organizes diverse physical scenarios within its latent space.

The visualization reveals that adjacent points within the t-SNE manifold correspond to deformation fields whose strain energy density functions share nearly identical profiles, regardless of the underlying material classification or the absolute magnitude of the strain energy. Furthermore, context representations derived from deformation fields with different geometries and loading modes can also converge to adjacent latent regions, provided their energy density functions share similar profiles.

These observations suggest that ICM fulfills our design expectations: by distilling vast datasets, it learns an intrinsic manifold primarily determined by the profiles of stress-strain relationships rather than other ad hoc factors, such as material classes, strain energy magnitudes, geometries, and loading conditions. By projecting complex physical context onto this intrinsic manifold, ICM realizes a shared mechanism of in-context inference, thereby driving the robust generalization previously demonstrated.

\subsection*{Train-time scaling behavior}

\begin{figure}
    \centering
    \includegraphics[width=1.0\textwidth]{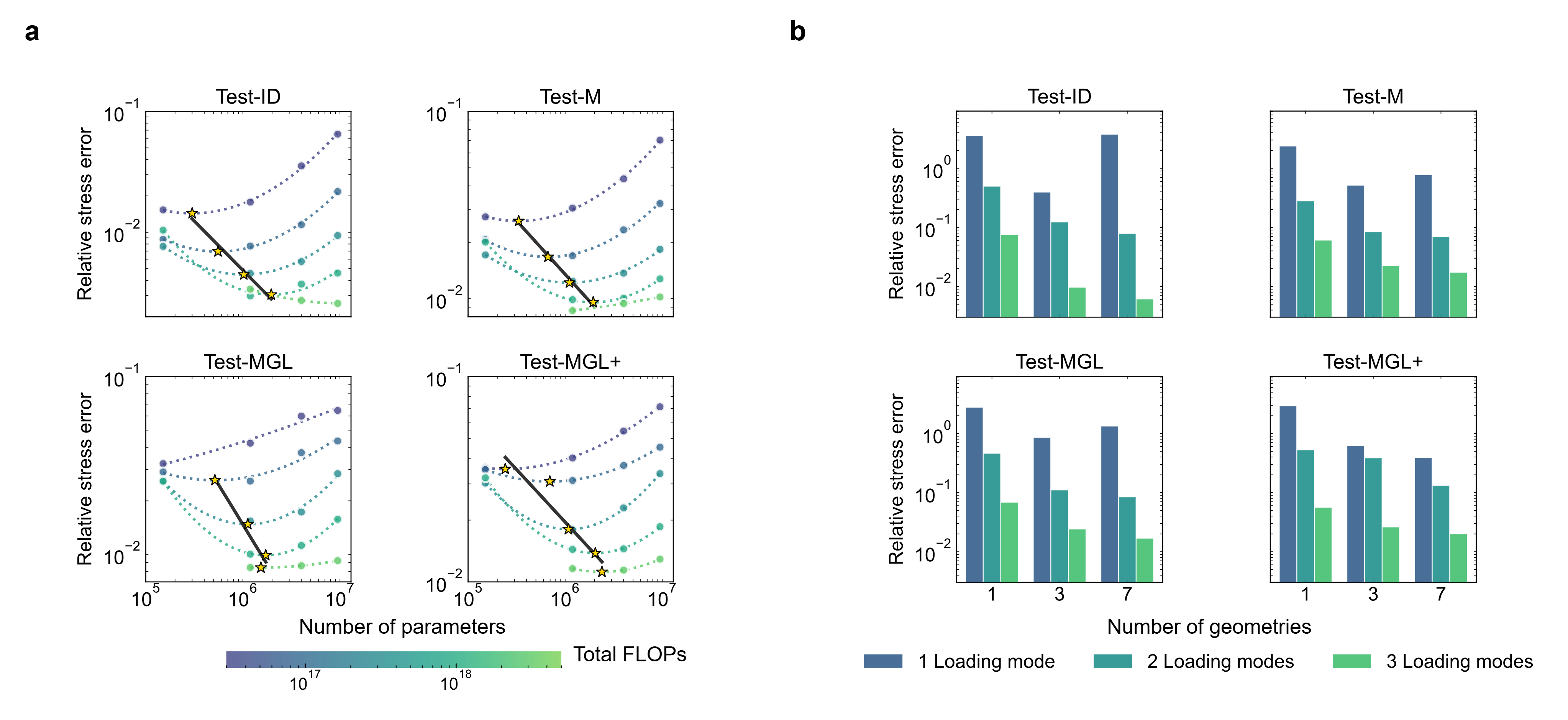}
    \caption{
    \textbf{Train-time scaling behavior of ICM over computational budget and data diversity.}
    \textbf{a,} Relative stress error as a function of model size (number of parameters) and computational budget across the four test sets. Each point represents a network configuration, colored by total floating-point operations (FLOPs). Dashed curves fit configurations sharing identical FLOPs via quadratic regression, and yellow stars mark the optimal model size that minimizes the error for each computational budget. The black lines fit these optima, illustrating that coupling greater computational budgets with appropriately larger architectures consistently drives down prediction errors. 
    \textbf{b,} Relative stress error evaluated as a function of training-data diversity on the same test sets. The horizontal axis denotes the number of training geometries (1 for (VII) only, 3 for (II)/(V)/(VII), and 7 for all training geometries), while the color indicates the number of loading modes (1, 2, or 3) included in the training set, with results for 1- and 2-mode configurations averaged across all possible combinatorial selections. The monotonic decay in error confirms that the model benefits from expanding the diversity of training set.
    }
    \label{fig:6}
\end{figure}

Having demonstrated the remarkable test-time scaling capabilities of ICM, we now turn to its train-time scaling behavior. In particular, we study the reduction in prediction error along two orthogonal dimensions: (i) training computational budget and (ii) training data diversity (\cref{fig:6}). To rigorously isolate these effects, we performed two complementary studies: one scaling model size and computational budget with a fixed dataset, and the other one scaling data diversity while fixing the model size and computational budget.

For the first study, we use the same training dataset as other sections, and train a family of ICM networks across a wide spectrum of parameter counts and training steps, quantifying the computational budget via total floating-point operations (FLOPs) (\cref{fig:6}a). We estimated the minimum testing error across various model sizes for a given computational budget, which yields an empirical scaling behavior: an increase in the computational budget drives a consistent reduction in prediction error, provided the model size is appropriately scaled up. 

For the second study, we fix the 4\,M model size and $1.0 \times 10^5$ training steps, and vary the diversity of the training set along two axes: the number of geometries and loading modes. As shown in \cref{fig:6}b, the relative stress error decays as data diversity increases. This shows ICM's remarkable capacity to harness data diversity, effectively translating rich training scenarios into performance gains.

Collectively, these results confirm that, at least within the scope of our experiments, ICM consistently benefits from increased training resources. By successfully leveraging broader data diversity and greater computational budgets to achieve lower prediction errors, the model demonstrates a scalable behavior that holds great promise for training future physical foundation models.

\section*{Discussion}

This work introduces In-Context Modeling (ICM) as a retrain-free paradigm for computational science. By recasting the identification of physical relationships as an in-context inference problem, ICM departs from conventional modeling-by-optimization approaches that encode system-specific behavior in fixed parameters and require repeated retraining. Instead, a single model dynamically infers governing relationships by conditioning on observational data as physical context, enabling adaptation across diverse materials, geometries, and loading conditions through a single forward pass. This shift from parameter-centric learning to context-driven inference directly addresses a central challenge in computational science: developing models that generalize efficiently across heterogeneous physical systems.

ICM's instant and robust generalization is built upon three foundational pillars. First, the paradigm of modeling from physical context equips the network with the inherent adaptivity to process diverse scenarios through a forward pass without retraining. Second, building upon this paradigm, our physics-informed tokenization, preprocessing, network architecture, and training scheme enable ICM to realize a shared mechanism of projecting complex physical context onto an intrinsic manifold. Third, ICM possesses favorable scaling behaviors at both the training and inference stages, effectively harnessing increasing computational budget and data diversity to achieve performance gains. The synergy of this in-context modeling paradigm, physics-informed techniques, and scaling behavior ultimately drives the generalization across unseen physical systems.

By shifting from ``modeling-by-optimization'' to ``modeling-from-context'', ICM addresses a central challenge in computational science: how to build models that instantly generalize across physical systems. While this study primarily focuses on hyperelastic modeling, ICM is built upon a common mathematical structure shared across disparate physical disciplines. Therefore, the methodology, encompassing both the paradigm and technical designs, is readily adaptable to other physically constrained disciplines, with its demonstrated scaling behavior providing the confidence needed for broader challenges. Even when the governing physical constraints deviate from the aforementioned affine structure, this paradigm and technical designs hold strong potential for successful application with minor modifications. Looking further ahead, we envision the possibility of a unified foundation model, where problems from diverse disciplines are treated with a common model under the ICM paradigm. We leave such explorations to future study.

\section*{Methods}
\subsection*{Derivation of modeling objective in hyperelasticity}\label{sec:mathematical-modeling}

In this section, we show how the equilibrium equations are discretized and reformulated into the physics-informed tokenization.

Let $\bm X\in\Omega\subset\mathbb{R}^2$ denote the material coordinates in the reference configuration and $\bm u(\bm X)$ the displacement field. The deformation gradient is $\bm F=\mathbb{I} +\nabla_{\! \bm X}\bm u$, where $\mathbb{I}$ is the identity tensor. For isotropic materials, the strain energy density $\psi$ depends on the invariants of the right Cauchy-Green tensor $\bm C=\bm F^\top\bm F$. In two dimensions, we consider
\begin{equation}
I_1=\mathrm{tr}\,\bm C,\qquad I_3=\det\bm C.
\end{equation}
Here, $I_2$ is omitted because it is fully determined by $I_1$ and $I_3$. We therefore denote the invariant vector by $\bm I=(I_1,I_3)$ with the index set defined as $\mathcal{M} = \{1,3\}$.

For a hyperelastic material, the second Piola-Kirchhoff stress $\bm S$ is given by
\begin{equation}\label{eq:second-PK}
  S_{lr}=2\,\frac{\partial \psi}{\partial C_{lr}}
        =2\,\frac{\partial I_m}{\partial C_{lr}}\,\frac{\partial \psi}{\partial I_m}.
\end{equation}
This relation shows that the stress can be expressed in terms of the gradient of the strain energy density with respect to the invariants.

To derive the discretization, we consider the weak formulation of the force equilibrium equations in the absence of body forces~\cite{de2012nonlinear,belytschko2014nonlinear}, given by
\begin{equation}\label{eq:weak-form}
    \int_{\Omega} \bm P : \nabla \bm v~\mathrm{d}V
    - \int_{\partial\Omega_t} \bar{\bm t}\cdot \bm v~\mathrm{d}S = 0, \quad  \forall \bm v\in\mathcal V,
\end{equation}
where $\Omega$ denotes the reference configuration of the domain with boundary decomposition $\partial\Omega=\partial\Omega_u\cup\partial\Omega_t$, and $\bm P$ is the first Piola-Kirchhoff stress tensor. The functional space $\mathcal V$ comprises sufficiently smooth test functions satisfying $\bm v=\bm0$ on the Dirichlet boundary $\partial\Omega_u$. Here, $\bar{\bm t}$ represents the traction on the Neumann boundary $\partial\Omega_t$. We assume displacement-controlled loading such that $\bar{\bm t}=\bm0$ on $\partial\Omega_t$, thus the second integral term vanishes.

Using the shape functions $\mathsf{N}^n$ associated with node $n$, we approximate the displacement field via linear interpolation
\begin{equation}
  \bm u(\bm X)=\sum_n \mathsf{N}^n(\bm X)\,\bm u^n,
\end{equation}
where $\bm u^n\in\mathbb{R}^2$ is the nodal displacement. Substituting this approximation into the weak form (\cref{eq:weak-form}) and choosing admissible test functions $\bm v$ yield the nodal force components
\begin{equation}\label{eq:nodal-force-continuous}
  f_i^{n}=\int_{\Omega} P_{ir}\,\frac{\partial \mathsf{N}^n}{\partial X_r}(\bm X)\,\mathrm{d}V
  \quad (i=1,2),
\end{equation}
where Einstein summation notation is adopted here and throughout the subsequent derivations. The discretized equilibrium equations can then be written as
\begin{equation}\label{eq:weak-f}
\begin{aligned}
\bm{f}^{n} &=\bm 0,&&\forall n \in \mathcal{N}^{\rm in}, &&& \text{(interior equilibrium)}\\
\sum_{n\in \mathcal{N}^{\rm b}_{j,k}} \bm{f}^{n} &= \bm F_{j,k}, &&\forall (j,k) \in \mathcal{B}, &&&  \text{(boundary equilibrium)}
\end{aligned}
\end{equation}
where $\bm{f}^{n} = (f_1^{n}, f_2^{n})$ is the nodal force, $\mathcal{N}^{\rm in}$ is the set of interior nodes, $\mathcal{N}^{\rm b}_{j,k}$ is the boundary node set for the $j$-th boundary condition in the $k$-th strain field, $\bm F_{j,k}$ is the corresponding external force, and $\mathcal{B}$ denotes the index set of all $(j, k)$ pairs.

Substituting $\bm P=\bm F\bm S$ and \cref{eq:second-PK} into \cref{eq:nodal-force-continuous} yields
\begin{equation}\label{eq:nodal-force-linear}
  f_i^{n}
  =\int_{\Omega} 2\,F_{il}\,\frac{\partial I_m}{\partial C_{lr}}\,
   \frac{\partial \psi}{\partial I_m}\,
   \frac{\partial \mathsf{N}^n}{\partial X_r}\,\mathrm{d}V,
\end{equation}
which is linear with respect to $\partial \psi/\partial I_m$.

We mesh $\Omega$ using $N_{\mathrm{e}}$ linear triangular elements. For linear triangles, $\nabla_{\!\bm X}\mathsf N^n$ and $\bm F$ are constant within each element, so the integral in \cref{eq:nodal-force-linear} can be evaluated exactly by element-wise quadrature
\begin{equation}\label{eq:nodal-force-discrete}
  f_i^{n}=\sum_{e=1}^{N_{\mathrm{e}}} w^eP^e_{ir}\frac{\partial \mathsf{N}^n}{\partial X^e_r}
         =\sum_{e=1}^{N_{\mathrm{e}}} 2w^eF^e_{il}\frac{\partial I^e_m}{\partial C^e_{lr}}
           \frac{\partial \psi}{\partial I^e_m}\frac{\partial \mathsf{N}^n}{\partial X^e_r},
\end{equation}
where the superscript $e$ denotes the element index and $w^e$ is the reference area of element $e$. We aggregate the coefficients of $\partial \psi / \partial I^e_m$ as
\begin{equation}\label{eq:A-coeff}
  A_{im}^{n,e}
  :=2w^eF^e_{il}\frac{\partial I^e_m}{\partial C^e_{lr}}
    \frac{\partial \mathsf{N}^n}{\partial X^e_r},
\end{equation}
and consequently the discrete nodal force can be written compactly as
\begin{equation}\label{eq:nodal-force-all}
  f_i^n=\sum_{e=1}^{N_{\mathrm{e}}}A_{im}^{n,e} \frac{\partial \psi}{\partial I_m^e},\quad i\in\{1,2\}.
\end{equation}
Since $\partial \mathsf{N}^n / \partial X^e_r \neq 0$ only for the elements adjacent to node $n$, we may equivalently restrict the sum to the neighboring elements as
\begin{equation}\label{eq:nodal-force-compact}
  f_i^n=\sum_{e \in \mathcal{E}(n)}A_{im}^{n,e} \frac{\partial \psi}{\partial I_m^e},\quad i\in\{1,2\},
\end{equation}
where $\mathcal{E}(n)$ is the set of elements connected to node $n$.

Combining \cref{eq:nodal-force-compact} with \cref{eq:weak-f} and denoting the matrix form of $A_{im}^{n,e}$ as $\bm A^{n,e}$ and the vector form of $I_m^e$ as $\bm I^e$, we recover the discrete equilibrium equations in the form of \cref{eq:equilibrium-internal,eq:equilibrium-external}
\begin{equation}
    \begin{aligned}
        \sum_{e\in\mathcal{E}(n)}\bm A^{n,e} \nabla \psi (\bm I^e) & =\bm 0, \quad \forall n \in \mathcal{N}^{\rm in},\\
        \sum_{n\in \mathcal{N}^{\rm b}_{j,k}}\sum_{e\in\mathcal{E}(n)}\bm A^{n,e}\nabla \psi (\bm I^e)&=\bm F_{j,k}, \quad \forall (j,k) \in \mathcal{B}.
    \end{aligned}
\end{equation}
\subsection*{Derivation of modeling objective in nonlinear diffusion}\label{sec:diffusion-modeling}

In this section, we present nonlinear diffusion as another example of the ICM paradigm, illustrating how the same methodology extends beyond hyperelasticity to a different physical setting through an appropriate physics-informed tokenization.

Let $\bm x\in\Omega\subset\mathbb{R}^2$ denote the spatial coordinate and $c(\bm x,t)$ the concentration field. We consider a diffusion process governed by Fick's law~\cite{bergman2017heatmass}
\begin{equation}
    \bm q = -\bm D(c)\nabla c,
    \label{eq:fick_law}
\end{equation}
where $\bm q$ is the diffusive flux and $\bm D(c)\in\mathbb{R}^{2\times 2}$ is the diffusivity tensor. In two dimensions, we consider
\begin{equation}
    \bm D(c) =
    \begin{bmatrix}
        D_{11}(c) & D_{12}(c) \\
        D_{12}(c) & D_{22}(c)
    \end{bmatrix}.
\end{equation}

To derive the discretized transport equations, we consider the weak formulation of the mass conservation equation in the absence of volumetric source terms, formulated as
\begin{equation}\label{eq:diffusion-weak}
    \int_{\Omega}
    v\,\frac{\partial c}{\partial t}\,\mathrm dV
    +
    \int_{\Omega}
    \nabla v\cdot\big(\bm D(c)\nabla c\big)\,\mathrm dV
    =
    0,
    \qquad \forall v\in\mathcal V,
\end{equation}
where $\Omega$ denotes the spatial domain, and the functional space $\mathcal V$ comprises sufficiently smooth scalar test functions satisfying $v=0$ on $\partial\Omega$, since Dirichlet boundary conditions are imposed on the entire boundary.

Using the shape functions $\mathsf N^n$ associated with node $n$, we approximate the concentration field at time $t^m$ by
\begin{equation}
    c^m(\bm x)=\sum_n \mathsf N^n(\bm x)\,c^{n,m},
\end{equation}
where $c^{n,m}$ denotes the nodal concentration at time $t^m$, and we write $c^m(\bm x):=c(\bm x,t^m)$. For each time step $m$, substituting the finite-element approximation into \cref{eq:diffusion-weak} and applying a backward-Euler discretization in time yield
\begin{equation}\label{eq:diffusion-nodal}
    \int_{\Omega}
    \mathsf N^n
    \frac{c^m-c^{m-1}}{\Delta t^m}\,\mathrm dV
    +
    \int_{\Omega}
    \nabla \mathsf N^n \cdot
    \big(
    \bm D(c^m)\nabla c^m
    \big)\,\mathrm dV
    =0,
    \qquad
    \forall n\in\mathcal N^{\rm in},
\end{equation}
where $\mathcal N^{\rm in}$ denotes the set of interior nodes.

We mesh $\Omega$ using $N_{\mathrm e}$ linear triangular elements. For each element $e$, both $\nabla \mathsf N^n$ and $\nabla c^m$ are constant. We further approximate the diffusivity tensor as constant within each element and evaluate it at the representative element concentration
\begin{equation}
    c^{e,m}:=\frac{1}{|\Omega^e|}\int_{\Omega^e} c^m(\bm x)\,\mathrm dV.
\end{equation}
For linear triangular elements, $c^{e,m}$ is equivalently the concentration at the element centroid. The nodal balance \cref{eq:diffusion-nodal} can thus be written as
\begin{equation}
    \sum_{e=1}^{N_{\mathrm e}} A_{ij}^{n,e,m} D_{ij}(c^{e,m})
    = \sum_{e=1}^{N_{\mathrm e}} b^{n,e,m},
\end{equation}
where
\begin{equation}
    A_{ij}^{n,e,m} := w^e
    \frac{\partial \mathsf N^n}{\partial x_i^e}
    \frac{\partial c^m}{\partial x_j^e},
\end{equation}
and
\begin{equation}
    b^{n,e,m}:=
    -\int_{\Omega^e}
    \mathsf N^n
    \frac{c^m-c^{m-1}}{\Delta t^m}\,\mathrm dV.
\end{equation}
Since $\mathsf N^n$ vanishes on elements not adjacent to node $n$, we may equivalently restrict the summation to the neighboring elements as
\begin{equation}\label{eq:diffusion-local}
    \sum_{e\in\mathcal E(n)}
    \bm A^{n,e,m} : \bm D(c^{e,m}) = b^{n,m},
    \qquad
    \forall n\in\mathcal N^{\rm in},\ \forall m\ge 1,
\end{equation}
where $\mathcal E(n)$ is the set of elements connected to node $n$, $\bm A^{n,e,m}$ denotes the matrix form of $A_{ij}^{n,e,m}$, $b^{n,m} = \sum_{e\in\mathcal E(n)} b^{n,e,m}$. This yields a local affine constraint on the constitutive response $\bm D(c)$, which serves as the modeling objective in nonlinear diffusion.
\subsection*{Network inputs and architecture}\label{sec:network-architecture}

The core of the ICM workflow is a network $g_{\bm\theta}$ defined in \cref{eq:g-theta}. Given a set of deformation tokens $\{(\bm A^{n,e}, \bm I^e)\}_{e\in \mathcal{E}(n)}$ as context, $g_{\bm\theta}$ predicts $\nabla \tilde{\psi} (\bm I)$~\cite{yang2023context,yang2024pde,cao2024vicon,yang2025fine}. The network consists of context-based normalizations, followed by an attention-based architecture. The detailed architecture is described below.

\paragraph{Context aggregation}
With interior equilibrium formulated by \cref{eq:equilibrium-internal}, we represent the local equilibrium state of an interior node $n$ by assembling the set $\{(\bm A^{n,e}, \bm I^e)\}_{e\in \mathcal{E}(n)}$ into a deformation token, with each pair $(\bm A^{n,e}, \bm I^e)$ treated as a deformation subtoken. This physics-informed tokenization yields a modular representation of the underlying stress-strain relationship, allowing tokens from different strain fields, geometries, and loading configurations of the same material to be flexibly assembled into a common physical context.

During training, the physical context for each material is assembled through a hierarchical sampling strategy. First, we construct a candidate pool of strain fields by uniformly sampling the number of involved geometries (1 to 7), loading modes (1 to 3), and loading magnitude within each loading mode (1 to 10) (\cref{fig:2}a). Next, 1 to 5 strain fields are randomly drawn from this customized pool. Finally, we uniformly sample 500 to 1500 deformation tokens from these selected fields to construct the physical context.

During testing, we aggregate all tokens from contextual deformation fields instead of applying random sampling. By retaining the entirety of the available tokens without any information loss, we ensure that ICM fully exploits the most comprehensive physical context possible, thereby maximizing inference accuracy and robustness.

\paragraph{Normalization of the coefficient matrices}

Discretized strain fields may involve non-uniform meshes, which can lead to large variations in the magnitude of the coefficient matrices $\bm A^{n,e}$, as $\bm A^{n,e}$ is sensitive to mesh resolution, an ad-hoc factor irrelevant to the underlying material response. To eliminate this non-physical sensitivity, we exploit the linear homogeneity of the interior equilibrium equations (\cref{eq:equilibrium-internal}) to normalize $\bm A^{n,e}$ in a node-wise and direction-wise manner. Specifically, for each node $n$ and spatial direction $i$, we define the normalized coefficients as
\begin{equation} \label{eq:A_norm}
\begin{aligned}
    &\bar{A}_{im}^{n,e} = A_{im}^{n,e} / \eta_i^n, \quad \forall e\in \mathcal{E}(n), \forall m \in \mathcal{M}, \\
    &\eta_i^n = \Big(\sum_{e' \in \mathcal{E}(n)} \sum_{m'} (A_{im'}^{n,e'})^2\Big)^{1/2},
\end{aligned}
\end{equation}
where $\eta_i^n$ is the local root-sum-square magnitude of the $i$-th force component at node $n$. Such normalization suppresses variations caused purely by discretization while preserving the underlying equilibrium presented to the network.

\paragraph{Normalization of the strain invariants}

To standardize the range of strain invariants in both contexts and queries, we first center all of them by the strain invariants at the undeformed state $\bm I_0=(2,1)^\top$, and then normalize them by the root-mean-square deviation from $\bm I_0$ calculated over $\mathcal{K}$, i.e., the collection of all node-element pairs $(n,e)$ in the context. This context-based normalization ensures that the input invariants are properly centered and adaptively scaled relative to the contextual deformation state.

\paragraph{Attention-based architecture}

Each deformation token is initially mapped to a context embedding in an independent manner. First, for each normalized deformation token, we map its subtokens to latent vectors via a shared linear projection. These vectors are then processed by position-encoding-free self-attention blocks. Finally, we apply average pooling to these vectors to produce a context embedding for each token. Such an embedding is permutation-invariant to subtoken ordering, in line with the physical interpretation that a node's equilibrium state is independent of the indexing of its neighboring elements.

Next, we facilitate information exchange within the context and enable the query to adaptively aggregate the context. In particular, for each normalized query, we first apply a linear projection to obtain a query embedding. The query and context embeddings are then processed by a stack of attention blocks. In each block, the query embedding attends to the context embeddings to extract the information relevant to the queried invariant state, while the context embeddings are updated by self-attention to capture interactions within the context, with two attention mechanisms sharing the same parameters for simplicity. Again, we apply no positional encoding here, making sure the processed query embeddings are invariant to the ordering of context embeddings. Moreover, since query embeddings do not attend to one another, multiple queries can be evaluated in parallel and independently. Finally, another linear layer is applied to the query embeddings to get the output of $g_{\bm\theta}$.

\subsection*{Training and inference}\label{sec:train-inference}
We train $g_{\bm\theta}$ by enforcing the interior equilibrium constraints in \cref{eq:equilibrium-internal}. At inference, a scaling factor $\alpha$ is applied to the network output to satisfy the boundary conditions in \cref{eq:equilibrium-external}.
 
\paragraph{Dimensionless loss function}
The predicted nodal force is the assembly of predicted element-wise contributions, i.e.,
\begin{equation}\label{eq:element_force}
\begin{aligned}
    \hat{\bm F}^n(\bm\theta) &= \sum_{e \in \mathcal{E}(n)} \hat{\bm F}^{n,e}(\bm\theta), \\
    \hat{\bm F}^{n,e}(\bm\theta) &= \bm A^{n,e} g_{\bm\theta}(\mathcal{C})(\bm I^e), 
\end{aligned}
\end{equation}
where $\hat{\bm F}^n(\bm\theta)$ is the predicted nodal force at node $n$, $\hat{\bm F}^{n,e}(\bm{\theta})$ represents the predicted force contribution from element $e$ to node $n$, and $\mathcal{E}(n)$ denotes the set of elements connected to node $n$.

Directly minimizing the norm of $\hat{\bm F}^n(\bm\theta)$ would lead to a trivial collapsing solution, as the network can simply suppress the overall amplitude of the output toward zero to minimize the objective. To circumvent this, we define a dimensionless loss function
\begin{equation}\label{eq:loss-fn}
    \mathcal{L}(\bm\theta) = \frac{
        |\mathcal{N}_{\text{in}}|^{-1} \sum_{n \in \mathcal{N}_{\text{in}}} \left\| \hat{\bm F}^n(\bm\theta) \right\|^2
    }{
        |\mathcal{K}|^{-1} \sum_{(n,e) \in \mathcal{K}} \left\| \hat{\bm F}^{n,e}(\bm\theta) \right\|^2
    }.
\end{equation}
In this formulation, the denominator acts as a self-normalizing factor derived from the context. This objective function is invariant to the scaling of the network output, forcing the network to focus on the physical equilibrium during training.

\paragraph{Post-scaling}
The interior equilibrium equations in \cref{eq:equilibrium-internal} are homogeneous in the energy gradient $\nabla \psi(\bm I)$. As a result, if $\nabla \psi(\bm I)$ satisfies the interior equilibrium equations, then so does $\alpha \nabla \psi(\bm I)$ for any scalar $\alpha$. In other words, the neural network prediction can only approximate $\nabla \psi(\bm I)$ up to a scaling factor. At inference, we determine such a scaling factor from the boundary conditions in \cref{eq:equilibrium-external}.

In particular, for the $j$-th boundary condition of the $k$-th contextual strain field, we obtain the predicted external force
\begin{equation}
    \hat{\bm F}_{j,k}
    =
    \sum_{n\in \mathcal N^{\mathrm b}_{j,k}}
    \hat{\bm F}^{n},
\end{equation}
where $\mathcal N^{\mathrm b}_{j,k}$ is the corresponding boundary node set. Since the external force is prescribed as a scalar magnitude along the loading direction, the comparison is performed on directional projections rather than Cartesian components. This avoids singular or unstable ratios that may arise when an individual Cartesian component is zero or nearly zero. Accordingly, we calculate the unscaled prediction of external force magnitude as
\begin{equation}
    \hat f_{j,k}
    =
    \hat{\bm F}_{j,k}\cdot \bm d_{j,k},
\end{equation}
where $\bm d_{j,k}$ is the loading direction. We also use $f_{j,k}$ to denote the corresponding true loading force magnitude calculated in the same way.

By comparing the unscaled prediction and true external forces across all available boundary conditions, a single scaling factor can be determined as
\begin{equation}
\begin{aligned}
    \alpha = & \left(
    \frac{1}{|\mathcal B|}
    \sum_{(j,k)\in\mathcal B}
    \alpha_{j,k}
    \right),\\
    & \alpha_{j,k}  = \frac{\hat f_{j,k}}{f_{j,k}}.
\end{aligned}
\end{equation}
The final prediction of the energy gradient is then given by
\begin{equation}
 \nabla \psi(\bm I) = 
    \alpha^{-1} \nabla \tilde{\psi}(\bm I)
    \approx
    \alpha^{-1} g_{\bm\theta}(\mathcal C)(\bm I).
\end{equation}

\paragraph{Evaluation metrics}
To assess the model performance quantitatively, we introduce the relative prediction error $(\cdot)^\mathrm{err}$ between the predicted and true physical quantities. For instance, to evaluate the stress prediction, we calculate
\begin{equation}\label{eq:relative-error-min-max}
    {S}^\mathrm{err} = \text{average} \left( \sqrt{\sum_{i,j=1}^2\left[
    \frac{S_{ij}^{\rm p} - S_{ij}^{\rm t}}
    {\max\left(S_{ij}^{\rm t}\right) - \min\left(S_{ij}^{\rm t}\right)}\right]^2} \right),
\end{equation}
where the average is taken over the whole mesh. Here, $S^{\rm p}_{ij}$ and $S^{\rm t}_{ij}$ $(i,j=1,2)$ are the components of the predicted and true stress tensors for a single element, respectively, and the denominator evaluates the range of each component $ S_{ij}^\mathrm{t}$ across the mesh.

\paragraph{Optimizer and scheduler}
We optimize the network using the Muon algorithm~\cite{liu2025muon}, which performs momentum SGD updates followed by Newton-Schulz orthogonalization on 2D parameter matrices to improve conditioning and generalization. Parameters that are not naturally represented as 2D matrices are optimized with AdamW~\cite{loshchilov2017decoupled}, providing a reliable fallback for remaining components.
For the learning rate, we employ a warmup-cosine schedule~\cite{goyal2017accurate,loshchilov2016sgdr}. The learning rate increases linearly from zero to a peak value of $5\times10^{-4}$ during the first $10\%$ of training steps, and then follows a cosine decay to $10\%$ of the peak value by the final step. This schedule mitigates early instabilities and yields smooth and stable convergence.

\section*{Code availability}

All code used for data generation, model training and analysis in this study is available via GitHub at https://github.com/scaling-group/icm-hyperelastic with an Apache license.

\bibliography{ref}

\section*{Acknowledgements}

The work is supported by the National Natural Science Foundation of China (Nos. 12132007 and T2488101). L.Y. acknowledges support from the National Research Foundation, Singapore, under the NRF fellowship (Project No. NRF-NRFF17-2025-0006). We acknowledge NUS IT’s Research Computing group for providing computational support. We would like to thank Prof. Qianxiao Li for the helpful discussions and insightful suggestions that improved the paper.

\section*{Author contributions}

H.G., C.C. and L.Y. contributed to the conception and design of the research. L.Y. led the method design. L.L., Z.L., S.L. and K.Z. performed the research. All authors contributed to writing the manuscript.

\section*{Competing interests}

The authors declare no competing interests.

\clearpage

\appendix

\section{Dataset}\label{sec:dataset}
\subsection{Strain energy models}

Our datasets cover various isotropic hyperelastic material classes. For isotropic hyperelasticity, the strain energy density function $\psi$ can be expressed as a scalar function of the three strain invariants of the right Cauchy-Green tensor $\bm C$, i.e., $\psi=\psi(I_1,I_2,I_3)$. In this work, we focus on plane strain. The detailed forms of $\psi$ used in this work are listed below.

\subsubsection{Polynomial model}\label{sec:dataset-polynomial}
The polynomial model is one of the most widely used phenomenological hyperelastic formulations. It defines the strain energy density function as a polynomial expansion of the strain invariants~\cite{rivlin1951large,hartmann2001numerical}. Owing to its flexible form, the model can reproduce a wide range of nonlinear stress-strain relationships through appropriate selection of the polynomial order and material coefficients. In this paper, we use polynomial models up to the sixth order. The strain energy density function is given by
\begin{equation}\label{eq:poly}
    \psi = \sum_{k=1}^6~\sum_{\substack{i,j \geq 0 \\ i+j = k}} 
C_{ij}(\bar{I}_1-3)^i(\bar{I}_2-3)^j
   + \sum_{m=1}^4 D_m (J-1)^{2m},
\end{equation}
where $\bar{I}_1=J^{-2/3}I_1$ and $\bar{I}_2=J^{-4/3}I_2$ are isochoric invariants, $J=\det\bm F=\sqrt{\det \bm C}$ is the volume ratio. The material parameters are the polynomial coefficients $C_{ij}$ and $D_m$ for all indices.

\subsubsection{Ogden model}
The Ogden model represents the strain energy density in terms of principal stretches, using a series of power-law terms~\cite{ogden1972large}. It is particularly effective in describing large deformations observed in rubber-like and biological tissues~\cite{kim2012comparison,budday2017mechanical}. We employ a six-term Ogden model, and the strain energy density function is given by
\begin{equation}\label{eq:ogden}
    \psi=\sum_{k=1}^6 \frac{2\mu_k}{\alpha_k^2}(\bar{\lambda}_1^{\alpha_k }+\bar{\lambda}_2^{\alpha_k }+\bar{\lambda}_3^{\alpha_k }-3)+\sum_{m=1}^4 D_m (J-1)^{2m},
\end{equation}
where $\lambda_1,\lambda_2,\lambda_3$ are the principal stretches, and we set the isochoric principal stretch $\bar{\lambda}_i=J^{-1/3}\lambda_i$ for each $i\in\{1,2,3\}$. The material parameters are $\mu_k$, $\alpha_k$, and $D_m$ for all indices.

\subsubsection{Pucci--Saccomandi model}
The Pucci--Saccomandi (PS) model refines the Gent model~\cite{gent1996new} by capturing the limiting-chain stretch of soft materials. It introduces an exponential-type formulation of the invariants, which leads to a logarithmic term in strain energy, and improves stability at large strains~\cite{pucci2002note}. In this work, we adopt its compressible form
\begin{equation}\label{eq:ps}
    \psi=-\frac{\mu J_{\rm{m}}}{2}\ln\left(1-\frac{\bar{I}_1-3}{J_{\rm{m}}}\right) 
    + C_2 \ln\left(\frac{\bar{I}_2}{3}\right) 
    + D\left(\frac{J^2-1}{2}-\ln J\right),
\end{equation}
where $\mu$ is the shear modulus and $J_{\rm{m}}$ is the limiting chain stretch. The material parameters are $C_2$ and $D$.

\subsubsection{Exp-ln model}
The exponential-logarithmic (Exp-ln) model combines exponential and logarithmic terms in the strain energy function. It features a simple mathematical form, physically meaningful parameters that can be determined from a single uniaxial test, and good agreement with experimental data~\cite{khajehsaeid2013hyperelastic}. In this work, we employ its compressible form as
\begin{equation}\label{eq:exp}
    \psi=\frac{\mu}{2}\left[\frac{1}{a}\exp[a(\bar{I}_1-3)]+b(\bar{I}_1-2)[1-\ln(\bar{I}_1-2)]-c\right]+D\left(\frac{J^2-1}{2}-\ln J\right).
\end{equation}
The material parameters are $a$, $b$, and $D$ with $c=a^{-1}+b$.

\subsubsection{Van der Waals model}
The van der Waals model incorporates concepts from statistical mechanics and molecular network theory~\cite{kilian1986use,marckmann2006comparison}. It captures the finite extensibility of polymer chains by introducing a singularity at the limiting stretch, thereby providing a physically motivated link between macroscopic mechanical behavior and molecular chain extensibility~\cite{marckmann2006comparison}. The strain energy density function is expressed as
\begin{equation}\label{eq:van}
    \psi=\mu\left[-(\lambda_{\rm{m}} ^2-3)[\ln(1-\eta)+\eta]-\frac{2a}{3}\left(\frac{\tilde{I}-3}{2}\right)^{3/2}\right]+D\left(\frac{J^2-1}{2}-\ln J\right),
\end{equation}
where $\tilde{I}=(1-\beta)\bar{I}_1+\beta\bar{I}_2$ and $\eta = \sqrt{(\tilde{I} - 3) / (\lambda_{\rm{m}}^2 - 3)}$. The material parameters are $a$, $\beta$, and $D$. 

\subsection{Data generation}
\subsubsection{Training set}\label{sec:trainset}
The training set is based on the polynomial model (\cref{eq:poly}) with 31 material parameters. In total, we generated 2,000 materials with various material parameters to construct the training set. More specifically, the training set can be divided into three subsets as follows.
\begin{enumerate}[label = (\Alph*)]
    \item For the first 500 materials, two randomly-selected coefficients of the lower-order ones $C_{ij}$ for $i+j=k\le3$ were sampled from a uniform distribution $\mU(0,100)$. The other $C_{ij}$ coefficients are set to zero. Two of the volumetric coefficients $(D_1,D_2,D_3,D_4)$ are randomly selected and sampled from $\mU(0,100)$, while the remaining two were set to zero. A post-sampling shift was applied by increasing both $C_{10}$ and $D_1$ by 1;
    \item For the next 500 materials, four randomly-selected $C_{ij}$ coefficients were sampled from a uniform distribution $\mU(0,100)$. The other $C_{ij}$ coefficients are set to zero. The same sampling strategy for the $D_m$ coefficients is applied as above.
    \item For the remaining 1,000 materials, the coefficients $C_{ij}$ and $D_m$ are sampled from $\mU(0,100)$ for all indices.
\end{enumerate}

To ensure numerical stability and comparable stress magnitudes among different models, the generated polynomial coefficients undergo a two-step normalization process as follows.

\begin{itemize}
    \item Basis-function standard deviation normalization: each basis function may contribute to the stress with a different magnitude. To balance their relative influence, we first compute the standard deviation of the stress response for each individual basis function under 1000 random 2×2 deformation gradients. The coefficients are then divided by these standard deviations, ensuring that all basis functions produce comparable stress amplitudes.
    
    \item Tangent stiffness normalization: even after the first step, different models can have very different initial stiffnesses. To standardize the small-strain response, we evaluate the stress difference under a small uniaxial compression and tension ($F_{11}=0.9$ and $F_{11}=1.1$ with fixed $F_{22}=1.0$) and estimate the effective tangent stiffness $\partial P_{11}/\partial F_{11}$. Each model’s coefficients are then divided by the stiffness so that all models have similar initial stiffness and stress levels.
\end{itemize}

\subsubsection{Test set}\label{sec:testset}

A total of four test sets, denoted as Test-ID (in distribution), Test-M (new materials), Test-MGL (new materials, geometries and loadings) and Test-MGL+ (new materials, geometries and extended loadings) are generated to comprehensively validate the model performance.

We generated the Test-ID set consisting of 400 materials with the polynomial model \cref{eq:poly}. The test set comprises three subsets of size $(100,100,200)$ and then is sampled with the same strategy as the training set.

The Test-M set evaluates out-of-distribution generalization by excluding polynomial models entirely. Instead, it consists of alternative hyperelastic formulations: the Ogden, Pucci--Saccomandi, Exp-ln, and van der Waals models. Specifically, for the Ogden model defined in \cref{eq:ogden}, we generated 200 parameter configurations divided into two complexity levels as below.

\begin{enumerate}[label = (\Alph*)]
    \item The first 100 materials are low-order models, where we set $\mu_k=0$ for $k\ge3$ and the remaining shear moduli $\mu_k$ were sampled from $\mU(1,101)$. To determine the exponent $\alpha_k$, we sampled $x\sim\mN(0,1)$ and set $\alpha_k$ as the clipping of $(2|x|+1)$ within the interval $[1.2,20]$. All volumetric coefficients $D$ were sampled from $\mU(0,100)$. A post-sampling shift was applied by increasing $D_1$ by 1 to avoid trivial incompressibility; 
    
    \item The remaining 100 materials are high-order models, where all the shear moduli $\mu_k$ were sampled from $\mU(1,101)$, $1\le k \le 6$. The exponents $\alpha_k$ were drawn similarly, but with a scaling factor of 10 instead of 2. The four volumetric coefficients $D_m$ were sampled with the same procedure, and the post-sampling shift also applies.
\end{enumerate}

To complete the Test-M set, 100 materials were generated with the material parameters randomly sampled from the ranges listed in \cref{tab:pucci,tab:exp_ln,tab:van} for the Pucci--Saccomandi, Exp-ln, and van der Waals models, respectively.

\renewcommand{\arraystretch}{1.2}
\begin{table}
    \centering
    \caption{Sampling distributions of the Pucci--Saccomandi model.}
    \begin{tabular}{lll}
    \toprule
    Parameter & Description & Distribution \\
    \midrule
    $\mu$ (MPa) & Shear modulus & $\mu\sim\mU(1, 101)$ \\
    $J_{\rm{m}}$ & Chain stretch limit & $\sqrt{J_{\rm{m}}}\sim\mU(4, 6)$ \\
    $C_2$ & Second invariant coefficient & $C_2\sim\mU(0, 100)$ \\
    $D$ (MPa) & Volumetric penalty coefficient & $D\sim\mU(1, 501)$ \\
    \bottomrule
    \end{tabular}
    \label{tab:pucci}
\end{table}
\renewcommand{\arraystretch}{1.0}

\renewcommand{\arraystretch}{1.2}
\begin{table}
    \centering
    \caption{Sampling distributions of the Exp-ln hyperelastic model.}
    \begin{tabular}{lll}
    \toprule
    Parameter & Description & Distribution \\
    \midrule
    $\mu$ (MPa) & Shear modulus & $\mu\sim\mU(1, 101)$ \\
    $a$ & Exponential coefficient & $a\sim\mU(0.1, 3.1)$ \\
    $b$ & Logarithmic coefficient & $b\sim\mU(0, 1)$ \\
    $D$ (MPa) & Volumetric penalty coefficient & $D\sim\mU(1, 501)$ \\
    \bottomrule
    \end{tabular}
    \label{tab:exp_ln}
\end{table}
\renewcommand{\arraystretch}{1.0}

\renewcommand{\arraystretch}{1.2} 
\begin{table}
    \centering
    \caption{Sampling distributions of the van der Waals model.}
    \begin{tabular}{lll}
    \toprule
    Parameter & Description & Distribution \\
    \midrule
    $\mu$ (MPa) & Shear modulus & $\mu\sim\mU(1, 101)$ \\
    $\lambda_{\rm{m}}$ & Chain stretch limit & $\lambda_{\rm{m}}\sim\mU(4, 6)$ \\
    $a$ & Nonlinearity coefficient & $a\sim\mU(0, 0.5)$ \\
    $\beta$ & Weighting factor between $I_1$ and $I_2$ & $\beta\sim\mU(0, 1)$ \\
    $D$ (MPa) & Volumetric penalty coefficient & $D\sim\mU(1, 501)$ \\
    \bottomrule
    \end{tabular}
    \label{tab:van}
\end{table}
\renewcommand{\arraystretch}{1.0} 

\section{Comparison with the ENN model}

The Equilibrium-based Neural Network (ENN) ~\cite{li2022equilibrium,li2025ennstressnet} provides a physics-informed baseline for learning a hyperelastic stress-strain relationship from full-field deformation measurements. Similar to the proposed ICM, the ENN leverages strain fields together with force equilibrium constraints and does not require stress labels. A key difference is that the ENN is trained for a single material using measured external forces, and thus does not transfer to unseen materials without retraining, whereas the ICM is designed for cross-material stress inference from contextual strain fields.

\subsection{ENN parameterization and invariance}
In the plane strain condition, the strain energy density function $\psi$ is parameterized as a scalar function of the invariants $(I_1,I_3)$. The ENN takes these two invariants as inputs and predicts $\psi_{\bm\theta}$ at each material point. The partial derivatives $\partial \psi/\partial I_m\;(m=1,3)$ are evaluated by automatic differentiation and are used to assemble the corresponding stresses and interior nodal forces. Material isotropy is embedded by expressing $\psi_{\bm\theta}$ in terms of invariants.

\subsection{Architecture modification}
The original ENN implementation uses a convolutional neural network (CNN) consisting of a stack of $1\times1$ convolutional layers to map strain component inputs to the stress component outputs. Since a $1\times1$ convolution performs an independent channel-wise affine transformation at each spatial location, it is equivalent to applying the same multilayer perceptron (MLP) to every material point. In our implementation, we therefore replace the $1\times1$ convolutional layers with an explicit MLP that takes the invariant pair $(I_1,I_3)$ as input and outputs the strain energy density $\psi_{\bm\theta}(I_1,I_3)$.

This change is purely notational and implementation-oriented, and does not change the functional form of the ENN model. It streamlines the evaluation of $\psi_{\bm\theta}(I_1,I_3)$ and its automatic-differentiation derivatives, and reduces architectural differences between ENN and the proposed ICM. The comparison thus primarily reflects differences in stress inference accuracy and deployment time.

\subsection{Equilibrium constraints and loss function}
The strain fields are discretized into finite elements to compute nodal forces $\bm f_n(\bm\theta)$. For interior nodes, force equilibrium requires $\bm f_n=\bm 0$. For boundary nodes, we enforce global equilibrium by matching the predicted boundary resultant forces to the experimentally measured external forces. Specifically, for each displacement-controlled boundary indexed by $i$, the predicted resultant force is computed by summing the nodal forces on that boundary as
\begin{equation}
    \hat{\bm F}_{b_i}(\bm\theta) = \sum_{n \in \mathcal{N}_{b_i}} \bm f_n(\bm\theta),
\end{equation}
and the prediction is compared with the measured resultant force $\bm F_{b_i}$.

The training loss of the ENN is
\begin{equation}\label{eq:ENN_loss}
    L = L_{\rm in} + w_{\rm b} L_{\rm b}
\end{equation}
as a weighted sum of the interior equilibrium loss $L_\mathrm{in}$ and the boundary loss $L_\mathrm{b}$ with a weight $w_{\rm b}$. The interior equilibrium loss $L_\mathrm{in}$ is defined as
\begin{equation}\label{eq:ENN_loss_free}
    L_{\rm in}= \frac{1}{N_{\rm free}} \sum_{n \in \mathcal{N}_{\rm free}} l\!\left(\bm f_n(\bm\theta), \bm 0\right),
\end{equation}
where $N_{\rm free}$ is the total number of interior nodes and $l(\cdot,\cdot)$ is the vector Huber loss evaluated by summing up the loss for each component as
\begin{equation}\label{eq:SmoothL1_vec}
    l(\bm x, \bm y) = \sum_k H(x_k, y_k),\qquad H(x,y)=
\begin{cases}
  \tfrac{1}{2}(x-y)^2, & \text{if } |x-y| < 1, \\
  |x-y| - \tfrac{1}{2}, & \text{if } |x-y| \ge 1.
\end{cases}
\end{equation}
The boundary loss $L_\mathrm{b}$ is defined by averaging over all displacement boundaries as
\begin{equation}\label{eq:ENN_loss_BC}
    L_{\rm b} = \frac{1}{N_{\rm b}} \sum_{i=1}^{N_{\rm b}} L_{b_i}, \quad
    L_{b_i} = l\!\left(\frac{\hat{\bm F}_{b_i}(\bm\theta)}{N_{b_i}}, \frac{\bm F_{b_i}}{N_{b_i}}\right),
\end{equation}
where $N_{\rm b}$ is the number of displacement boundaries and $N_{b_i}$ is the number of nodes on boundary $i$. The normalization by $N_{b_i}$ prevents the boundary term from being sensitive to the boundary node count.

\subsection{Normalization and force scaling}
The inputs $(I_1, I_3)$ are normalized using the mean and standard deviation computed from all training data points. To improve numerical stability, we also scale the nodal forces by a characteristic force magnitude estimated from the training set. Specifically, we define a force scale $s_f$ as the average of the absolute values of the measured external forces across all training samples, and divide all predicted nodal forces and boundary resultants by $s_f$ when evaluating \cref{eq:ENN_loss_free,eq:ENN_loss_BC}, which yields a dimensionless training objective and reduces sensitivity to the overall force magnitude.

\subsection{Model configurations and training details}
We trained the ENN models with a set of strain fields and measured boundary external forces by minimizing $L$ in \cref{eq:ENN_loss}, and three different model configurations summarized in \cref{table:ENN_model_size} were considered.
The boundary loss weighting coefficient $w_{\rm b}$ was fixed at $10^{-1}$. We trained the neural networks using the Adam optimizer with an initial learning rate of $10^{-3}$. We applied a StepLR learning-rate schedule, reducing the learning rate by a factor of $0.95$ every $100$ optimization steps.
\begin{table}[h]
\caption{The ENN model configurations used for comparison}
\label{table:ENN_model_size}
\begin{tabular}{cccc}
\toprule
Model  & Layers & Hidden size & Params \\ 
\midrule
Small  & 2      & 256         & 132K   \\
Medium & 8      & 768         & 4.14M  \\
Large  & 8      & 1024        & 8.40M  \\ 
\bottomrule
\end{tabular}
\end{table}

\section{Experimental setup}

In addition to the plane strain setting used throughout this work, we conducted an auxiliary training experiment under plane stress. The ICM training and the hyperparameter settings are the same as those for the plane strain case, and the only difference was that the stress-strain relationship is constrained by the plane stress condition. This is the only experiment in this study that adopts the plane stress assumption.

\subsection{Materials}

Two hyperelastic materials with different stiffnesses were used in the experiments. For clarity, we refer to them as the soft and hard materials. All specimens were fabricated by Wenext Factory (China) and characterized by large reversible deformations owing to their high elasticity.

\subsection{Specimen geometry and DIC measurements}

\begin{figure}
    \centering
    \includegraphics[width=0.75\textwidth]{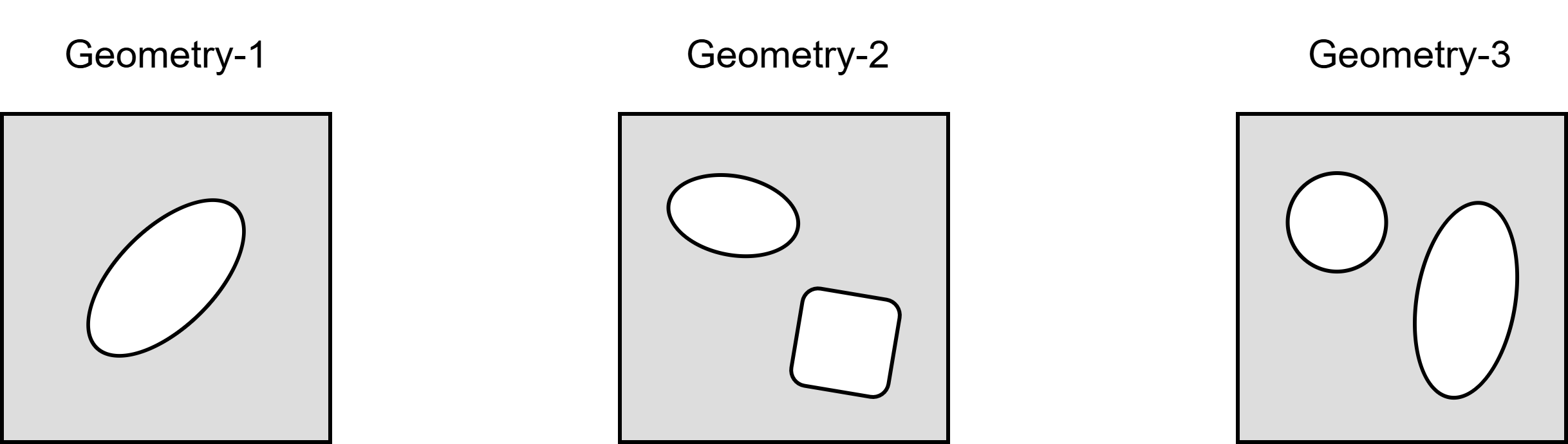}
    \caption{Three specimen geometries used in the experiments.} 
    \label{fig:exp-geometry}
\end{figure}

We designed three distinct specimen geometries, including configurations with geometric imperfections, to induce non-uniform deformation fields and thereby increase the diversity of the contextual strain fields (\cref{fig:exp-geometry}). For each test, full-field deformation data were obtained and subsequently processed into deformation tokens that serve as the context input for model validation. All specimens were subjected to uniaxial tensile loading, during which full-field deformation data were captured using a digital image correlation system. A random black-and-white speckle pattern was sprayed on the specimen surfaces to ensure optical tracking. Images were recorded at two frames per second, while the specimens were loaded at a displacement rate of 22.5 mm/min up to a total displacement of 40 mm. This setup enabled accurate measurement of displacement and strain fields across the entire specimen surface.

The acquired full-field deformation data allowed the ICM to infer the thickness-averaged stress in the specimen under the following assumptions. We denote the three-dimensional stress--strain relationship by $\bm P=\bm P(\bm F)$, where both the first Piola--Kirchhoff stress $\bm P$ and deformation gradient tensor $\bm F$ are three-dimensional second-order tensors. Because the specimens are thin relative to their in-plane dimensions and both free surfaces are traction-free, we adopted a plane stress approximation and assumed that the stress and deformation vary weakly through the thickness. Under the assumption, the out-of-plane shear components of both stress and strain vanish, and the out-of-plane normal stress is zero. Accordingly, $\bm F$ and $\bm P$ take the forms

\begin{equation}
    \left[\bm F\right] = \left[\begin{array}{ccc}
         F_{xX} & F_{yX} & 0\\
         F_{xY} & F_{yY} & 0\\
         0 & 0 & \lambda_3
    \end{array}\right], \quad
    \left[\bm P\right] = \left[\begin{array}{ccc}
         P_{xX} & P_{yX} & 0\\
         P_{xY} & P_{yY} & 0\\
         0 & 0 & 0
    \end{array}\right],
\end{equation}
where $\lambda_3$ represents the stretch in the thickness direction. The surface deformation measurements determine the in-plane components of $\bm F$, denoted as $\bm F_{\text{2D}}$, and we define $\bm P_{\text{2D}}$ as

\begin{equation}
    \left[\bm F_\text{2D}\right] = \left[\begin{array}{cc}
         F_{xX} & F_{yX}\\
         F_{xY} & F_{yY}
    \end{array}\right],\quad
    \left[\bm P_\text{2D}\right] = \left[\begin{array}{cc}
         P_{xX} & P_{yX}\\
         P_{xY} & P_{yY}
    \end{array}\right].
\end{equation}
The proposed ICM model learns the two-dimensional stress--strain relationship $\bm P_{\text{2D}}=\bm P_{\text{2D}}(\bm F_{\text{2D}})$, which can be interpreted as the three-dimensional stress--strain relationship $\bm P=\bm P(\bm F)$ constrained by the plane stress condition $P_{zZ}(\bm F_\text{2D},\lambda_3)=0$. Under the assumption of uniform through-thickness deformation, we can average the three-dimensional equation of stress equilibrium over the thickness direction to yield the in-plane equilibrium

\begin{align}
    \label{eq:experiment_2D_equilibrium}
        \frac{\partial P_{xX} }{\partial X} +\frac{\partial P_{xY} }{\partial Y }=0 \\
        \frac{\partial P_{yX} }{\partial X}+\frac{\partial P_{yY} }{\partial Y }=0
\end{align}
Based on $\bm F_{\mathrm{2D}}$ and $\bm P_{\mathrm{2D}}$, nodal forces can be computed in the same way as in the simulation setting, and the nodal force equilibrium condition used for training remains applicable. Moreover, we used two invariants derived from the in-plane deformation, denoted as $I_1^{\mathrm{2D}}$ and $I_3^{\mathrm{2D}}$, as the ICM input. Here, $I_1^{\mathrm{2D}}$ and $I_3^{\mathrm{2D}}$ are computed from $\bm F_{\mathrm{2D}}$. Under plane stress, only two invariants are independent, and the remaining invariant can be derived from $I_1^{\mathrm{2D}}$ and $I_3^{\mathrm{2D}}$. 

The total forces recorded by the uniaxial tension machine were divided by the initial thicknesses and assigned to the boundary total forces during the ICM inference. Although the normal strain and stress through the thickness were not directly measured, they are not required for training or inferring the in-plane stress under the plane stress assumption.

\subsection{Preprocessing of experimental data}

DIC provides the surface coordinates and displacements of tracked points on a regular, pixel-style grid. We denoised and interpolated the measured displacement field onto a set of triangular meshes using a radial basis function (RBF) interpolator~\cite{virtanen2020scipy} with a thin-plate-spline kernel and a linear polynomial tail (degree 1). The target meshes resembled the meshes used during training of the ICM model. We computed a characteristic point spacing $h$ from the DIC points as the median distance to each point's 7th nearest neighbor (computed in the 2D coordinate space). The smoothing parameter of the interpolator was chosen as $0.05 h^2$. This resulted in good interpolation in well-sampled interior regions, suppressed measurement noise and stable extrapolation where the domain was not fully covered by DIC data (e.g., the boundaries). For computational efficiency, the interpolation at each mesh node used a fixed-size local neighborhood (here 100 nearest DIC points).

To ensure reliable inference, we identified mesh nodes where the interpolated displacements are likely to be inaccurate due to insufficient local DIC support. Specifically, we computed a reference neighborhood radius $h_n$ as the median distance to the $n_e=8$ nearest neighbor among the DIC points. A mesh node was marked reliable if at least $n_e/2=4$ DIC points were located within a radius $h_n$ of that node; otherwise it was marked unreliable. During ICM inference, deformation tokens were constructed only for the reliable nodes and used as the context.

\subsection{Consistency of experimental validation}

\begin{figure}
    \centering
    \includegraphics[width=1.0\textwidth]{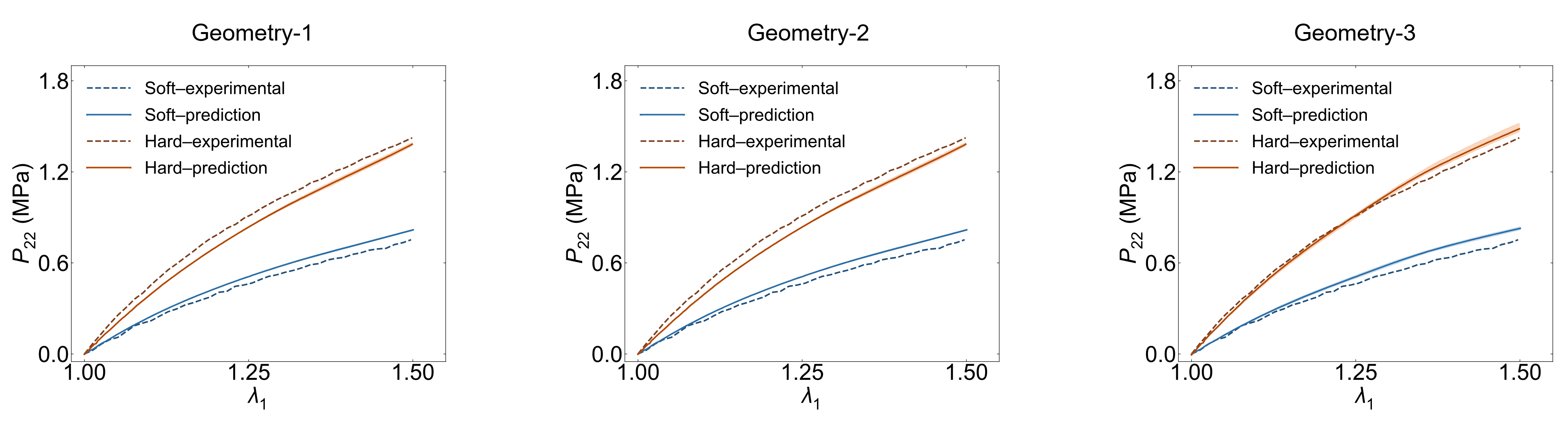}
    \caption{Predicted uniaxial nominal stress responses using the measured full-field deformation data of three geometries as context.} 
    \label{fig:exp-result}
\end{figure}

To assess the sensitivity of ICM inference to the choice of context geometry, we performed an additional validation study in which the measured full-field deformation data from each specimen geometry were separately converted into deformation tokens and used as the context input to predict the uniaxial tensile response (\cref{fig:exp-result}). The predicted nominal stress responses obtained from these three contexts are highly consistent, indicating that the model provides stable stress predictions despite specimen-to-specimen variations and differences in the geometry used to construct the context.

\section{t-SNE embedding}

In the main text, we visualized the high-dimensional context embeddings produced by the ICM for samples drawn from the four test sets using a two-dimensional t-SNE projection. Despite distribution shifts in material class, geometry, and loading mode, the embeddings consistently cluster deformation states with similar mechanical response patterns into nearby regions, revealing cross-material response equivalences. Here we provide the explicit material models, including the strain energy density form and the corresponding parameters, for the representative points highlighted in Fig.5 of the main text. The detailed configurations are summarized in \cref{table:t-snematerial} for reference.

\clearpage

\begin{table}[ht]
\centering
\renewcommand{\arraystretch}{1.2} 
\caption{Representative material models corresponding to the highlighted t-SNE points.}
\label{table:t-snematerial}
\begin{tabular}{lll >{\raggedright\arraybackslash}p{0.3\linewidth} p{0.45\linewidth}}

\toprule
No. & Mat. & Geo. & Loading & Parameters \\
\midrule

1 & VdW & II & shear \newline $\bar{u}_1=0,\ \bar{u}_2/L=0.15$ &
$\begin{aligned}[t]
\mu &= 32.3797,&\lambda_{\mathrm m}&=4.3952, \\
a &=0.0596,&\beta&=0.4010, \\
D &=45.0786&&
\end{aligned}$\\
\addlinespace

2 & PS & II & shear \newline $\bar{u}_1=0,\ \bar{u}_2/L=0.15$ &
$\begin{aligned}[t]
\mu &=45.5183,&\lambda_\mathrm {m}&=32.1880, \\
C_2 &=91.1765,&D&=104.2352
\end{aligned}$\\
\midrule

3 & EXp-Ln & VI & biaxial \newline $\bar{u}_1/L=0.5,\ \bar{u}_2/L=0.125$ &
$\begin{aligned}[t]
\mu &=5.4041\times10^{1},&a &=3.0775\times10^{-1},\\
b&=9.3254\times10^{-1},&D &=4.4402\times10^{2}
\end{aligned}$\\
\addlinespace

4 & VdW & IV & biaxial \newline $\bar{u}_1/L=0.5,\ \bar{u}_2/L=0.5$ &
$\begin{aligned}[t]
\mu &=18.6790,&\lambda_\mathrm {m}&=4.2527, \\
a &=4.7719\times 10^{-1},&\beta&=2.0768\times10^{-1}, \\
D &=3.6567\times10^{2}
\end{aligned}$\\
\addlinespace

5 & PS & IV & biaxial \newline $\bar{u}_1/L=0.5,\ \bar{u}_2/L=0.3125$ &
$\begin{aligned}[t]
\mu &=6.5509,&\lambda_\mathrm {m}&=17.1022,\\
C_2 &=6.3923,&D&=339.7878
\end{aligned}$\\
\midrule

6 & Ogden & X & proportional biaxial (2-side) \newline $\bar{u}_1/L=0.18,\ \bar{u}_2/L=0.09$&
$\begin{aligned}[t]
\mu_1 &=11.6752,&\mu_2&=15.7796, \\
\alpha_1 &= 4.6862,&\alpha_2&=2.6835, \\
D_1 &=54.3825,&D_2&=58.4697, \\
D_3 &=75.5499,&D_4&=29.5771
\end{aligned}$\\
\addlinespace

7 & VdW & X & equal biaxial \newline $\bar{u}_1/L=0.14,\ \bar{u}_2/L=0.14$ &
$\begin{aligned}[t]
\mu &=66.5179,&\lambda_\mathrm {m}&=4.6732, \\
a &=2.1922\times 10^{-1},&\beta&=4.5983\times10^{-1}, \\
D &=1.1403\times10^{2}
\end{aligned}$\\
\midrule

8 & Exp--Ln & IV & uniaxial \newline $\bar{u}_1/L=0.3,\ \bar{u}_2=0$ &
$\begin{aligned}[t]
\mu &=40.7137,&a&=3.0398, \\
b &=0.4611,&D&=22.5349
\end{aligned}$\\
\addlinespace

9 & Poly. & IV & uniaxial \newline $\bar{u}_1/L=0.2,\ \bar{u}_2=0$ &
$\begin{aligned}[t]
C_{10} &=0.1656,&C_{20}&=2.4958,\\
C_{02} &=1.3797,&C_{04}&=0.2843,\\
C_{42} &=0.1132,&D_1&=0.1324, \\
D_2 &=0.1212,&D_4&=0.0306
\end{aligned}$\\
\addlinespace

10 & Ogden & III & uniaxial \newline $\bar{u}_1/L=0.3,\ \bar{u}_2=0$ &
$\begin{aligned}[t]
\mu_1 &=88.1378,&\mu_2&=91.6428, \\
\mu_3 &=82.8531,&\mu_4&=73.1291, \\
\mu_5 &=11.3618,&\mu_6&=98.2382, \\
\alpha_1 &=13.4329,&\alpha_2&=17.0617, \\
\alpha_3 &=16.3020,&\alpha_4&=7.9690,\\
\alpha_5 &=8.9464,&\alpha_6&=2.1765,\\
D_1 &=62.4539,&D_2&=72.2618, \\
D_3 &=73.2496,&D_4&=13.8722
\end{aligned}$\\

\bottomrule
\multicolumn{5}{p{\linewidth}}{\footnotesize \textbf{Note:} The parameters $\mu$, $C$, and $D$ are in MPa. The cases are numbered clockwise starting from the top left in Figure 5 of the main text.}
\end{tabular}
\end{table}

\end{document}